\documentclass[aps,prd,reprint,groupedaddress]{revtex4-1}

\usepackage{natbib}
\usepackage[pdftex]{graphicx}
\usepackage{color}
\usepackage{amsmath}
\usepackage{rotating}
\usepackage{wasysym}

\providecommand{\prob}{\mathrm{P}}   % probability
\providecommand{\resi}{r}                    % residual term
\providecommand{\tosubia}{{o}_{\mathrm{i}1}}     % input 1
\providecommand{\tosubib}{{o}_{\mathrm{i}\Delta}}   % input 12
\providecommand{\tosubxa}{{o}_{\mathrm{x}1}}   % input 1
\providecommand{\tosubxb}{{o}_{\mathrm{x}\Delta}} % input 12

\providecommand{\speca}{S_{\mathrm{x}1}}
\providecommand{\specb}{S_{\mathrm{x}\Delta}}
\providecommand{\specab}{S_{\mathrm{x}1:\mathrm{x}\Delta}}

\begin{document}
\title{Bayesian parameter estimation in the second LISA Pathfinder Mock Data Challenge}
% Autors
\author{M Nofrarias}
\author{C R\"over}
\author{M Hewitson}
\author{A Monsky}
\author{G Heinzel}
\author{K Danzmann}
\affiliation{Max-Planck-Institut f\"ur Gravitationsphysik (Albert-Einstein-Institut) and 
Leibniz Universit\"at Hannover, 30167~Hannover, Germany}
\author{L Ferraioli}
\author{M Hueller}
\author{S Vitale}
\affiliation{Dipartimento di Fisica, Universit\`a di Trento, and I.N.F.N., Gruppo di Trento, 38050 Povo, Italy}
% Date
\date{\today}

\begin{abstract}
A main scientific output of the LISA Pathfinder mission is to provide
a noise model that can be extended to the future gravitational wave
observatory, LISA\@. The success of the mission depends thus upon a
deep understanding of the instrument, especially the ability to
correctly determine the parameters of the underlying noise model.  In
this work we estimate the parameters of a simplified model of the LISA
Technology Package (LTP) instrument. We describe the LTP by means of a
closed-loop model that is used to generate the data, both injected
signals and noise. Then, parameters are estimated using a Bayesian
framework and it is shown that this method reaches the optimal
attainable error, the Cram\'{e}r-Rao bound. We also address an important
issue for the mission: how to efficiently combine the results of
different experiments to obtain a unique set of parameters describing
the instrument.
\end{abstract}

%\keywords{}

\maketitle
%\tableofcontents

\newcommand{\martin}[1]{\textcolor{blue}{\textit{Martin: #1}}}
\newcommand{\miquel}[1]{\textcolor{red}{\textit{Miquel: #1}}}
\newcommand{\christian}[1]{\textcolor{green}{\textit{Ch.: #1}}}

\newcommand{\etc}{\textit{etc}}
\newcommand{\ie}{\textit{i.e.}}
\newcommand{\da}{\textsc{LTPDA}}
\newcommand{\code}[1]{\texttt{#1}}
\newcommand{\tbd}[0]{\textcolor{red}{TBD}}
\newcommand{\ltpda}[0]{\textsc{LTPDA}}

\newcommand{\note}[1]{\textcolor{blue}{(#1)}}

\section{Introduction}

LISA Pathfinder~\cite{Armano09} is an ESA mission, with some NASA
contributions, that aims at testing key technologies for the future
space gravitational wave observatory, LISA~\cite{Bender00}.  The main
aim is to demonstrate the ability to put a test mass in to free-fall
at a level of $3\times10^{-14}\,\rm{ms}^{-2}/\sqrt{Hz}$ at 1\,mHz.
The LISA Technology Package (LTP) is the main instrument on-board LISA
Pathfinder. It comprises two test masses enclosed in inertial sensors
which are in turn housed inside individual vacuum tanks, composing the
so called
Gravitational Reference Sensor~\cite{Dolesi03}. The two tanks
are then mounted to a support structure which also holds an optical
bench between the tanks. The optical bench and the associated
interferometry are part of the Optical Metrology
System~\cite{Heinzel05}.  In order to reach the goal stated above, the
full LTP must be characterized and optimized. This will involve
developing a full parametric noise model of the instrument, which will
be improved over the course of the mission.

The LISA Pathfinder mission comprises a series of experiments. Many of
the experiments aim to reduce the noise in the system so as to produce
the quietest residual acceleration measurement possible. Other
experiments will aim to characterize the instrument. This
typically involves determining the various parameters that go into
the physical model of the instrument. Clearly, a good model is needed
to be able to target and reduce particular noise sources, whereas
reducing the various noise sources leads to a more sensitive instrument.
Various experiments will be repeated under different conditions, and
as the noise is reduced, we would expect that the determination of the
physical parameters will become more and more accurate. One essential
aspect of this multiple-experiment mission is the ability to include
the results from analyzing the previous experiments in further
experiments, and in particular, it will be necessary to combine the
various experiments to gain the best knowledge about the particular
physical parameters. The analysis procedures and software
need therefore to remain flexible in order to react to the results of the
experiments as they are performed.  This paper presents a Bayesian
analysis for determining particular physical parameters of the system.
Using a Bayesian framework leads to a natural way of combining a
series of experiments.  The result of one analysis becomes prior
information in subsequent analyses.  The analysis is presented for a
reduced set of physical parameters in the context of the Mock Data
Challenges (MDC)~\cite{Hewitson09} that are being carried out during
the development of the data analysis procedures for the mission. In
MDC1~\cite{Monsky09} the focus was on developing a simple model of the
system, together with establishing routines for calibrating the
measured test mass displacements back to equivalent residual external
test mass accelerations. In MDC2, the focus shifts to parameter
estimation. The analysis and procedures presented in this paper
represent one of the methods being developed for the mission.

% $Id: model.tex,v 1.15 2010/05/06 09:19:56 miquel Exp $

\section{The second LTP Mock Data Challenge}
\begin{figure*}[t]
  \begin{center}
     \includegraphics[width = 1\columnwidth]{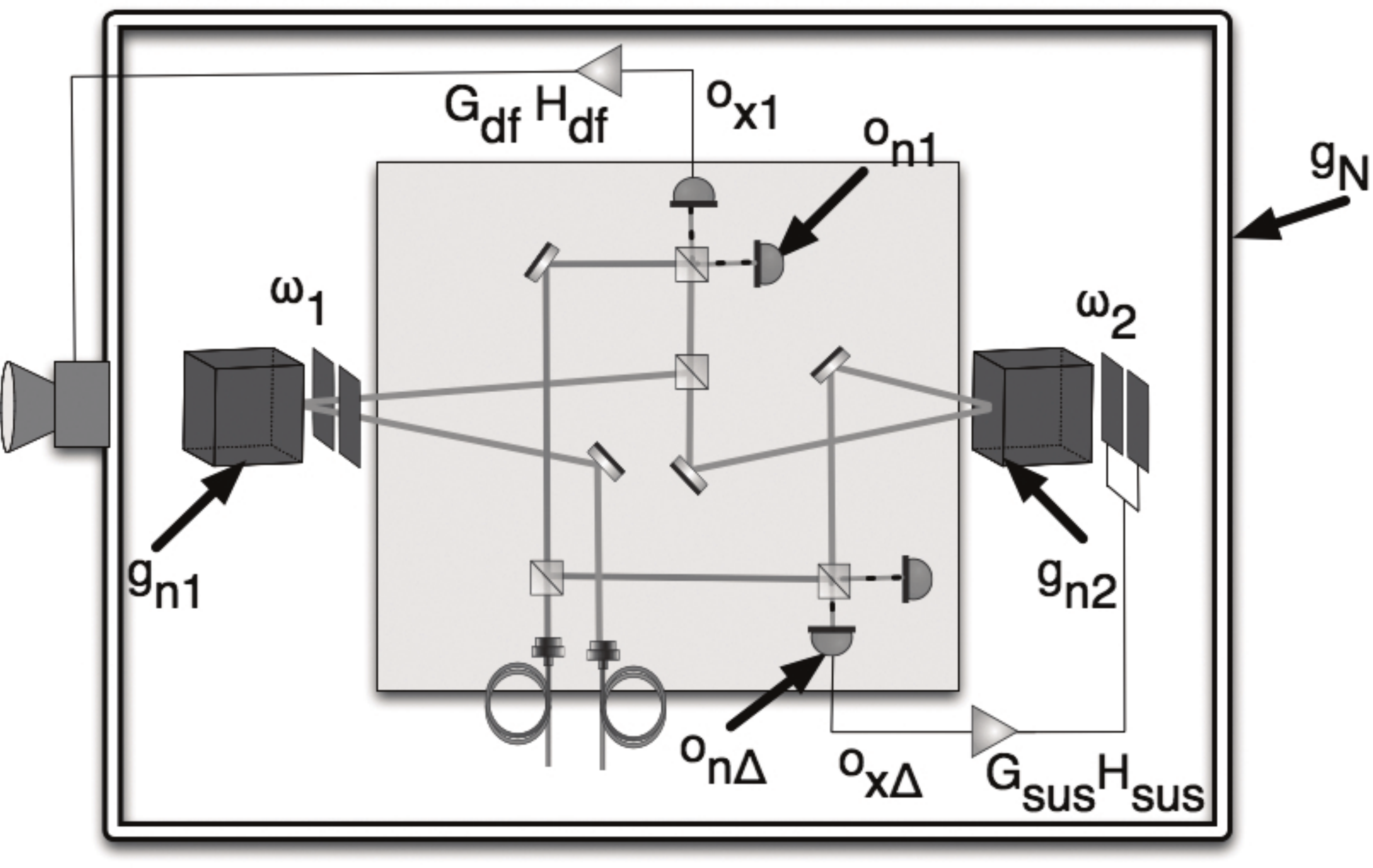}   %1
    \includegraphics[width = 1\columnwidth]{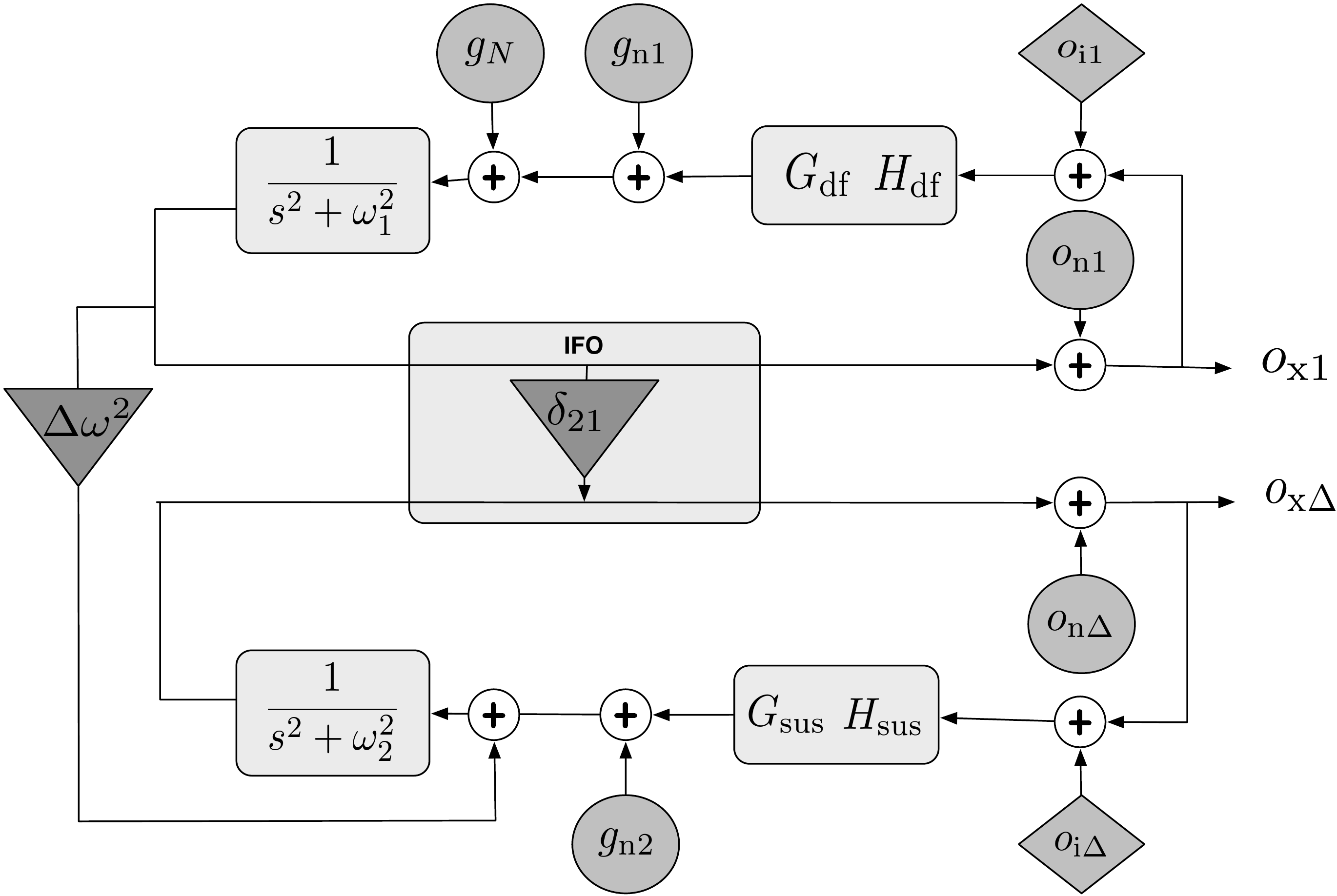}         %1
    \caption{\label{fig.scheme} The LTP MDC2 model. \emph{Left:} 
    Simplified scheme of the LTP instrument. Only two out of the four 
    heterodyne interferometers are represented here: the one measuring spacecraft to first 
    test mass distance, $o_{x1}$, and the one measuring test mass to test mass 
    distance, $o_{\Delta x}$. See text for a description of terms appearing in the picture.
    \emph{Right:} The previous is described as a control loop:  
     the boxes describe the
	interferometer (IFO), controllers and dynamics of the test
	masses. The circles represent noise contributions, diamonds
	are signal injection points and the triangles denote
	cross-couplings between the first ($o_{x1}$) and second
	channel ($o_{\Delta x}$).}
  \end{center}
\end{figure*}

The aim of the second MDC was to develop and test reliable methods to
accurately estimate the parameters of the LTP noise model during
flight operations. In order to focus on methods and not on model
complexity, it was decided to keep a very similar model as the one
analyzed during the first MDC\@. The basic difference regarding the
previous challenge is that now 5 parameters are considered as
degrees-of-freedom of the system, which need to be determined by
stimulating the system using injected signals.  It is worth recalling
that the first MDC did not include any signal injection in the data,
since it was designed as a test of the calibration of displacement
noise to acceleration noise, and therefore only a noise measurement
(signal free) was simulated. The current challenge is therefore a
natural extension to the first one.

It is important to notice that, due to the nature of the LISA
Pathfinder mission, our description of the system necessarily needs to
deal with the closed-loop dynamics of the spacecraft and test masses
together.

The description that we show in Section~\ref{sec.model} is therefore a
closed-loop system where we take into account the feedback between
different components and show where parameters and noise contributions
enter in the non-linear model that is described in terms of transfer
functions in the frequency domain. We want to recall that this
approach differs from the one used to model LISA to the date. The data
generators that are providing data in the LISA Mock Data Challenges
\cite{Rubbo04,Vallisneri05,Petiteau08} are focused on the geometry of
the spacecraft configuration, since the main concern is, in that case,
the suppression of frequency noise due to the unequal arms. But, on
the other hand, they consider additive noise sources inside each
spacecraft. A second important remark is that LISA generators model
noise sources as white gaussian contributions. This is clearly
unrealistic and could be particularly misleading in the relevant
region around 1\,mHz, since each of the noise sources will contribute
with a $f^{-p}$ ($p \simeq 1$) power spectrum that will set the low
frequency performance of LISA\@. LTP is designed to study that
region and therefore our model needs to describe these low frequency
contributions in more detail. The noise models and the parameters used
are described in Section~\ref{sec.params}. The
description provided in this paper will complement the one already
existing within the LISA community and will facilitate the interaction
between both communities to a common goal, which is a realistic
understanding of the LISA instrument.

In terms of implementation, it is worth mentioning that the current
challenge is completely implemented as \mbox{LTPDA}
tools~\cite{Hewitson09}, which means that any user of this tool has the
means available to produce LTP-like data (as described in the
following section) by executing a relatively simple
MATLAB~\cite{MATLAB} script.

%---------------------------------------------------------------------

\subsection{Dynamical model \label{sec.model}}

When compared with other space missions, the LTP is a very flexible
instrument in terms of the possible operational scenarios. It can be
configured to use different combinations of the available sensors
onboard, either optical or capacitive, with the aim of performing
different geodesic measurements, or even to work as an
accelerometer. The aim of the second MDC was not to cover all of these
possibilities but to analyze the instrument behavior for a fixed
operating mode: the main science mode ---~described as
M3 mode in \cite{Fichter05}. Moreover, this
control scheme is reduced in this analysis to the one-dimensional case
in order to simplify the model and focus on the analysis. In this
simplified model, the $x$ position of both test masses is controlled
by means of the optical readouts.  A first interferometer measures the
relative distance between test mass 1 and the spacecraft,
$\mathrm{x}_1$.  This is a relatively noisy measurement since the
noise of the spacecraft's micro-Newton thrusters appears directly in
the measurement. A second interferometer measures the relative
distance between both free falling test masses. This channel, that we
call $\mathrm{x}_{\Delta}$ in the rest of this paper, will be the one
giving an unprecedently quiet measurement of the 
differential acceleration (or displacement)
between two test masses, since the contribution of the thruster noise
effectively cancels out~\cite{Bortoluzzi04}.

The model of the LTP dynamics control loop is shown in
Figure~\ref{fig.scheme}. The right panel of this figure shows two control loops for the
two measurement channels that we just described: $\mathrm{x}_1$ and
$\mathrm{x}_{\Delta}$.
This schematic representation of the closed loop system can be
analytically expressed in terms of the following set of
equations~\cite{TN3045}
\begin{eqnarray}
{\mathbf{D}}\cdot \vec{q} & = & \vec{g},  \label{eq.dyn1is} \nonumber\\
\vec{g} & = & - {\mathbf{C}}\cdot (\vec{o}+ \vec{o}_i) - \vec{g_n},  \label{eq.dyn2is}\\
\vec{o} & = & {\mathbf{S}}\cdot \vec{q} + \vec{o_n}, \label{eq.dyn3is} \nonumber
\end{eqnarray}
where $\mathbf{D}$ is the dynamical matrix, $\mathbf{C}$ is the
controller, and $\mathbf{S}$ stands for the sensing matrix 
(the interferometer in our case), i.e., the
matrix translating the position of a test mass, $\vec{q}$, into the
interferometer readout, $\vec{o}$. Subindex~$n$ stands for noise
quantities, either sensing noise ($\vec o_n$) or force noise ($\vec
g_n$) and subindex~$i$ stands for the injected signals ($\vec
o_i$). All of these are 2-dimensional vectors with components referring to 
the  $\mathrm{x}_1$ and $\mathrm{x}_{\Delta}$ channels respectively,
\begin{eqnarray}
  \vec{q}  = \left(
  \begin{array}{c}
   x_1 \\ 
   x_{\Delta}  
  \end{array}   \right) & , & 
    \vec{o}  = \left(
  \begin{array}{c}
   o_{x1} \\ 
    o_{x \Delta} 
  \end{array}   \right), \nonumber \\
  \vec{o}_i  = \left(
  \begin{array}{c}
   o_{i1} \\ 
   o_{i\Delta}  
  \end{array}   \right) & , &
  \vec{o}_n  = \left(
  \begin{array}{c}
   o_{n1} \\ 
   o_{n\Delta}  
  \end{array}   \right), \\
    \vec{g}_n & = & \left(
  \begin{array}{c}
   g_{n1} - g_N \\ 
   g_{n2} - g_{n1}  
  \end{array}   \right) . \nonumber
\end{eqnarray}
The last equation shows how any noisy force applied to the spacecraft ($g_N$) 
is only measured in the first channel (if there were no cross terms). On the other hand,
the differential channel is sensitive to the difference of force noise applied to 
the first and the second test mass, $g_1$ and $g_2$ respectively.

The matrices read as
%\begin{widetext}
\begin{eqnarray}
  \mathbf{D}  & = & \left(
  \begin{array}{cc}
  s^2 + \omega^2_{1} & 0 \\ 
  \omega^2_{2} -\omega^2_{1} &  s^2 + \omega^2_{2}
  \end{array}   \right), \nonumber \\ \nonumber \\
  \mathbf{C} & = & \left(
  \begin{array}{cc}
   G_{\rm df} \, H_{\rm df} & 0\\
   0 &  G_{\rm sus} \, H_{\rm sus}
  \end{array}
  \right),  \label{eq.dyn} \\ \nonumber \\
  \mathbf{S} & = & \left(
  \begin{array}{cc}
  1 & 0\\
   \delta_{21} & 1
  \end{array}
  \right), \nonumber
\end{eqnarray}
%\end{widetext}
where $\omega_{1}$ and $\omega_{2}$ are the stiffness ---
the steady force gradient across the test mass housing 
per unit mass~\cite{Bortoluzzi04}  ---
coupling the motion of each test mass to the motion of the spacecraft;
$G_{\rm df}$ and $G_{\rm sus}$ are constant factors acting as
calibration factors of the controller, $H_{\rm df}$ and $H_{\rm sus}$. 
These are the control laws of the loop and will be 
considered known transfer functions in the following; 
$\delta_{21}$ is the
interferometer cross-coupling, a small term accounting for the
imperfection of the interferometer that will produce a spurious signal
in the differential channel when only the first test mass moves.
The interferometer has no coupling 
going from $o_{\Delta}$ to $o_{1}$ 
and therefore we set  $\delta_{12} = 0$ in the sensing matrix. 
The previous are the 5~parameters that we will consider in the following
discussion, the ones characterizing the dynamics of the
instrument.

The leading diagonal terms in Equation~(\ref{eq.dyn}) describe the
dynamics of each channel (for example, $s^2 + \omega^2_{1}$ is
Newton's law in the Laplace domain for the first test mass, with
$\omega_{1}$ being the test mass stiffness), and the control law (for
example, $G_{\rm df}\, H_{\rm df}$ stands for the drag-free transfer
function controller on the first test mass, multiplied by a constant
calibration factor, $G_{\rm df}$). The off-diagonal terms are the
cross-couplings between the two channels appearing as triangles in
Figure~\ref{fig.scheme}. From Equation~(\ref{eq.dyn}) we can compute
the response of the interferometer once all the dynamical and noise
parameters are given as
\begin{eqnarray}
  \vec{o}  & = &  ({\mathbf{D}}\cdot{\mathbf{S}}^{-1}+{\mathbf{C}})^{-1} (-
{\mathbf{C}}\,\vec{o}_i + \vec{g}_n + {\mathbf{D}}\cdot {\mathbf{S}}^{-1} \vec{o}_n).
\end{eqnarray}
This equation describes the interferometer output
and will be the variable that we will use to evaluate the
interferometer response. It may be useful to express the nominal
output as a signal and two noise terms:
\begin{equation}
  \vec{o}  =  \mathbf{G_s}(\Theta)\,\vec{o}_i + 
\mathbf{G_{no}}(\Theta)\,\vec{o}_n  +  \mathbf{G_{ng}}(\Theta )\,\vec{g}_n,
  \label{eq.oo_all}
\end{equation}
where $\Theta = \lbrace G_{\rm df}, G_{\rm df}, \omega^2_1,
\omega^2_2, \delta_{21}\rbrace$ are the unknown model parameters we are
interested in determining.
\begin{table}[t]
\caption{Parameters for the LTP MDC2 model \label{tbl.model}}
\begin{ruledtabular}
\begin{tabular}{lccc}
 \multicolumn{4}{c}{\textsc{Dynamical Parameters}}\\
\hline
Parameter & \multicolumn{3}{c}{Value} \\
\hline
$\rm G_{df}$     & \multicolumn{3}{c}{0.8}  \\
$\rm G_{sus}$  & \multicolumn{3}{c}{1.15}  \\
$\rm \omega^2_{1}$   & \multicolumn{3}{c}{$-11 \times 10^{-7}$}  \\
$\rm \omega^2_{2}$  & \multicolumn{3}{c}{$-22 \times 10^{-7}$}   \\
$\rm \delta_{21}$      &  \multicolumn{3}{c}{$1.35 \times 10^{-4}$}  \\
\hline
 \multicolumn{4}{c}{\textsc{Noise Parameters}}\\
\hline
Parameter & $o_{n1}/o_{n \Delta}$ & $g_{n1}/g_{n2}$ & $g_N$ \\
\hline
$p_1$ & $3.6\times 10^{-12}$  & $7\times 10^{-15} $ &  $2.5\times 10^{-10}$  \\
$p_2$ & $10\times 10^{-3}$     & $ 5\times 10^{-3}$   &  $12\times 10^{-3}$ \\
$p_3$  & 4.2                             & 3                              &  3.8 \\
$p_4$  & $1.8\times 10^{-3}$   & $4\times 10^{-4}$    &  $1\times 10^{-3}$\\
$p_5$ & 8                                 &  8                             & 8 \\
\end{tabular}
\end{ruledtabular}
\end{table}

Our model can be thought of as a first term which filters the input
signal~($\vec{o}_i$) and two further terms which filter the noise. It
must be stated that, since our final aim is to characterize the noise
model, the noise terms also contain information about our
parameters. But, since we will be working in a high signal-to-noise
ratio (SNR) regime, we will not consider this dependence in our
analysis and we will further simplify the model with the
approximations $\mathbf{G_{no}}(\omega, \Theta) \approx
\mathbf{G_{no}}(\omega) $ and $\mathbf{G_{ng}}(\omega, \Theta
) \approx \mathbf{G_{ng}}(\omega )$.  This allows us to rewrite
Equation~(\ref{eq.oo_all}) as
\begin{equation}
  \vec{o}  =  \mathbf{G_s}(\Theta )\,\vec{o}_i +  \vec{n},
  \label{eq.oo_all_simpl}
\end{equation}
where $\vec{n}$ now represents the overall noise of the
instrument. The first term then contains all the model dependence that
we will be able to test with our experiments.  The transfer function
in this formulation now has the following components
\begin{eqnarray}
G^{11}_s  & = & \frac{G_{\rm df}\, H_{\rm df}(\omega)}{\omega^2_{1} - \omega^2 + G_{\rm df}\,H_{\rm df}(\omega)}, \\
G^{12}_s  & = & 0, \\
G^{21}_s  & = & \frac{G_{\rm df} H_{\rm df} \left(\omega^2_2 -\omega^2_{1} +\delta_{21}  \left(\omega ^2-\omega^2_2 \right)\right)} {\left( \omega^2_1-\omega^2+ G_{\rm df} H_{\rm df} \right) \left(\omega^2_2-\omega^2+ G_{\rm lfs} H_{\rm lfs}  \right)} \label{eq.xcoupl},\\
G^{22}_s  & = & \frac{G_{\rm lfs}\, H_{\rm lfs}(\omega)}{\omega^2_{2} - \omega^2 + G_{\rm lfs}\,H_{\rm lfs}(\omega)}, 
\end{eqnarray}
where we can see that by injecting and measuring in the same
channel (\ie, testing the diagonal terms), we are able to determine
either $\lbrace G_{\rm df}, \omega^2_{1}\rbrace$ or $\lbrace G_{\rm
sus},\omega^2_{2} \rbrace$, and it is through the non-diagonal
(cross-coupling) term that we can determine the $\delta_{21}$
parameter and the difference between stiffnesses, $
\omega^2_2-\omega_1^2$.  The experiments in this MDC were designed to
test these possible combinations of injected signals, as described in
the following.

%---------------------------------------------------------------------
\subsection{Model parameters \label{sec.params} }

Our model is defined by a total of 30 parameters, which can be divided
into two groups: noise parameters and dynamical parameters. The first
ones are those ones used to set the noise shapes of the individual
noise contributions --- force noise~$\vec{g}_n$ and interferometer
read-out noise~$\vec{o}_n$ in Equation~(\ref{eq.oo_all}) --- that
will set the final instrument noise level. Each contribution is
described as
\begin{equation}
	S(\omega) = p^2_1 \left( 1 + \frac{1}{\left( \frac{\omega}{2\,\pi p_2}
\right)^{p_3}}+ \frac{1}{\left( \frac{\omega}{2\,\pi p_4} \right)^{p_5} }\right)^{1/2},
\end{equation}
and therefore 5~parameters are required for each of them, for a total
of 25 to describe all noise contributions. We need to add to these the
5 parameters that characterize the joint dynamical behavior of the
spacecraft and test masses. Only the latter will be the parameters
that we will be interested in recovering from the data in this
challenge. As stated above these are: stiffness for each test mass
($\rm \omega^2_{1},\omega^2_{2}$), calibration for each controller
($G_{\mathrm{df}}, G_{\mathrm{sus}}$) and interferometer
cross-coupling ($\delta_{21}$).

Table~\ref{tbl.model} contains all numerical values used in the second
Mock Data Challenge, and therefore fully characterizes the
model. Although the model allows for different noise levels for $x_1$
and $x_{\Delta}$ interferometer noise, we did not use this degree of
freedom and set both interferometers to behave equally. The same
applies to the force noise acting on both test masses.

%\begin{table*}[t]
%\caption{Parameters for MDC2 model \label{tbl.model}}
%\begin{ruledtabular}
%\begin{tabular}{lccccc}
%\multicolumn{6}{c}{Dynamical parameters} \\
%\hline
%Parameter & Value\\
%\hline
%$\rm G_{df}$     & \multicolumn{5}{c}{0.8}  \\
%$\rm G_{sus}$  & \multicolumn{5}{c}{1.15}  \\
%$\rm \omega^2_{1}$   & \multicolumn{5}{c}{$-11 \times 10^{-7}$}  \\
%$\rm \omega^2_{2}$  & \multicolumn{5}{c}{$-22 \times 10^{-7}$}   \\
%$\rm \delta_{21}$      &  \multicolumn{5}{c}{$1.35 \times 10^{-4}$}  \\
%\hline
%\multicolumn{6}{c}{Noise parameters} \\
%\hline
%$o_{n1}, o_{n2}$ & $3.6\times 10^{-12}$ & $10\times 10^{-3}$ & 4.2 & $1.8\times 10^{-3}$ & 8\\ 
%$g_{n1}, g_{n2}$ & $7\times 10^{-15}, 5\times 10^{-3}, 3, 4\times 10^{-4}, 9$ \\ 
%$g_N$       & $2.5\times 10^{-10}, 12\times 10^{-3}, 3.8,  1\times 10^{-3} , 8$ \\ 
%\end{tabular}
%\end{ruledtabular}
%\end{table*}

% $Id: experiments.tex,v 1.17 2010/05/06 08:32:13 miquel Exp $

\subsection{Experiments}

Three experiments were proposed for MDC2. These were originally
motivated by first studies about the sensitivity attainable by
injected signals during the mission~\cite{TN3045} and correspond to a
frequency sweep in the measurement bandwidth at four different
frequencies. Our experiments in MDC2 consider only the possibility of
injected signals as simulated interferometric signals, the so-called
\textsl{interferometric bias}, which we have labelled in
Equation~(\ref{eq.oo_all}) and in Figure~\ref{fig.scheme} as
$\vec{o}_i$. LISA Pathfinder will allow other kinds of injected
signals, for instance, forces applied to the spacecraft via the
thrusters or forces directly applied to the test masses via the
capacitive sensors but, as stated above, it is not the aim of this
work to explore all capabilities of the mission. In that sense,
extending the analysis to include all possible injection signals is
one of the aims of the forthcoming LISA Pathfinder MDCs. The three
proposed experiments for this challenge were the following:
\begin{figure}[t]
  \begin{center}
  \includegraphics[width = 1\columnwidth]{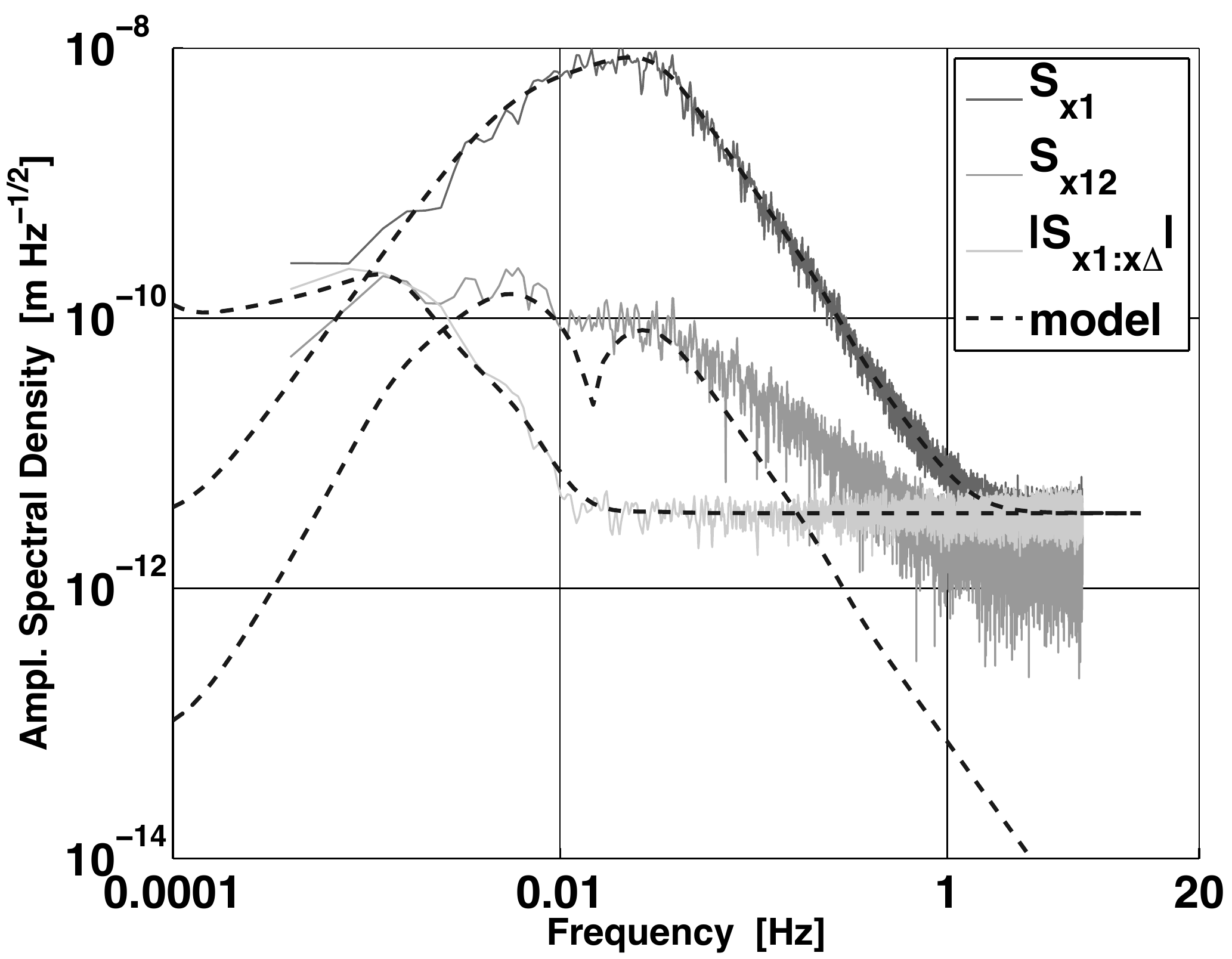}
  \caption{Amplitude spectral density of a noise realization of the LTP MDC2 
  noise model compared to analytical curves. We compare the noise of the
  first channel ($S_{\rm x1}$), the second channel ($S_{\rm x\Delta}$) 
  and the absolute value of the cross-spectra between both  ($S_{\rm x1:x\Delta}$) .
  \label{fig.noiseASD} }
\end{center}
\end{figure}

\begin{figure*}[t]
\begin{center}
        \line(1,0){200}\; \textbf{Experiment 1}\; \line(1,0){200} \\[0.2cm]
        \includegraphics[width = 0.75\columnwidth]{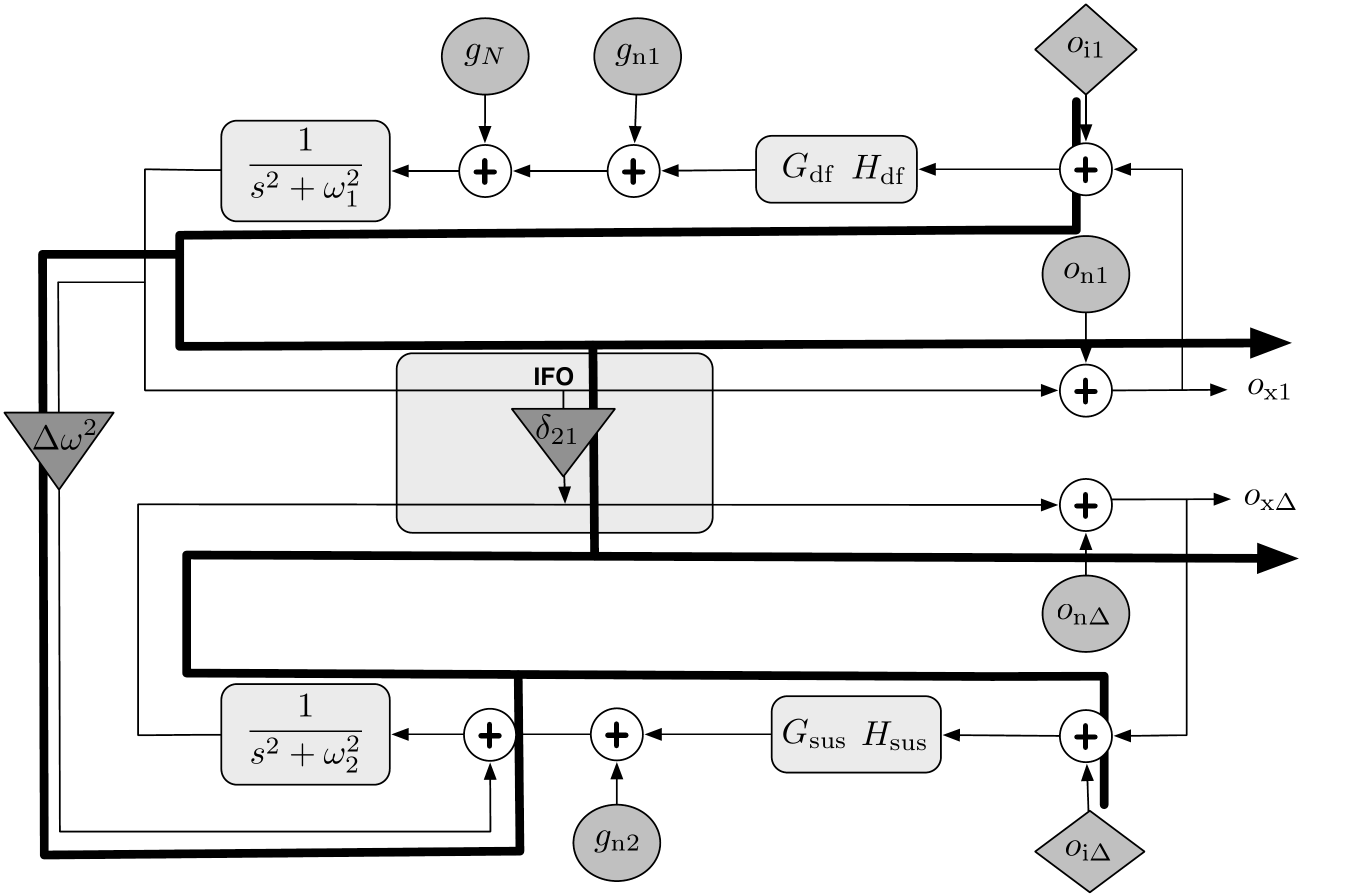}
        \includegraphics[width = 0.6\columnwidth]{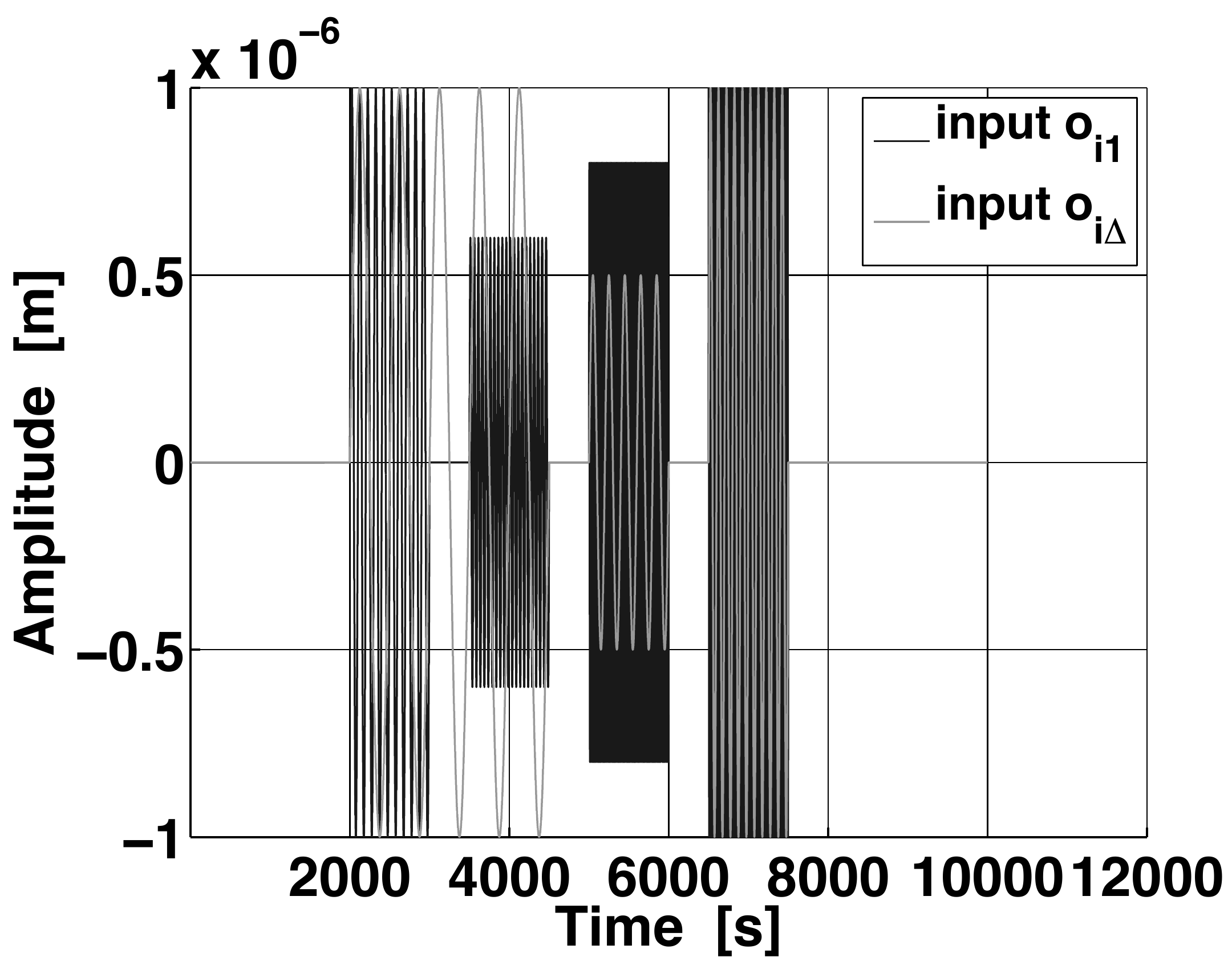}
        \includegraphics[width = 0.6\columnwidth]{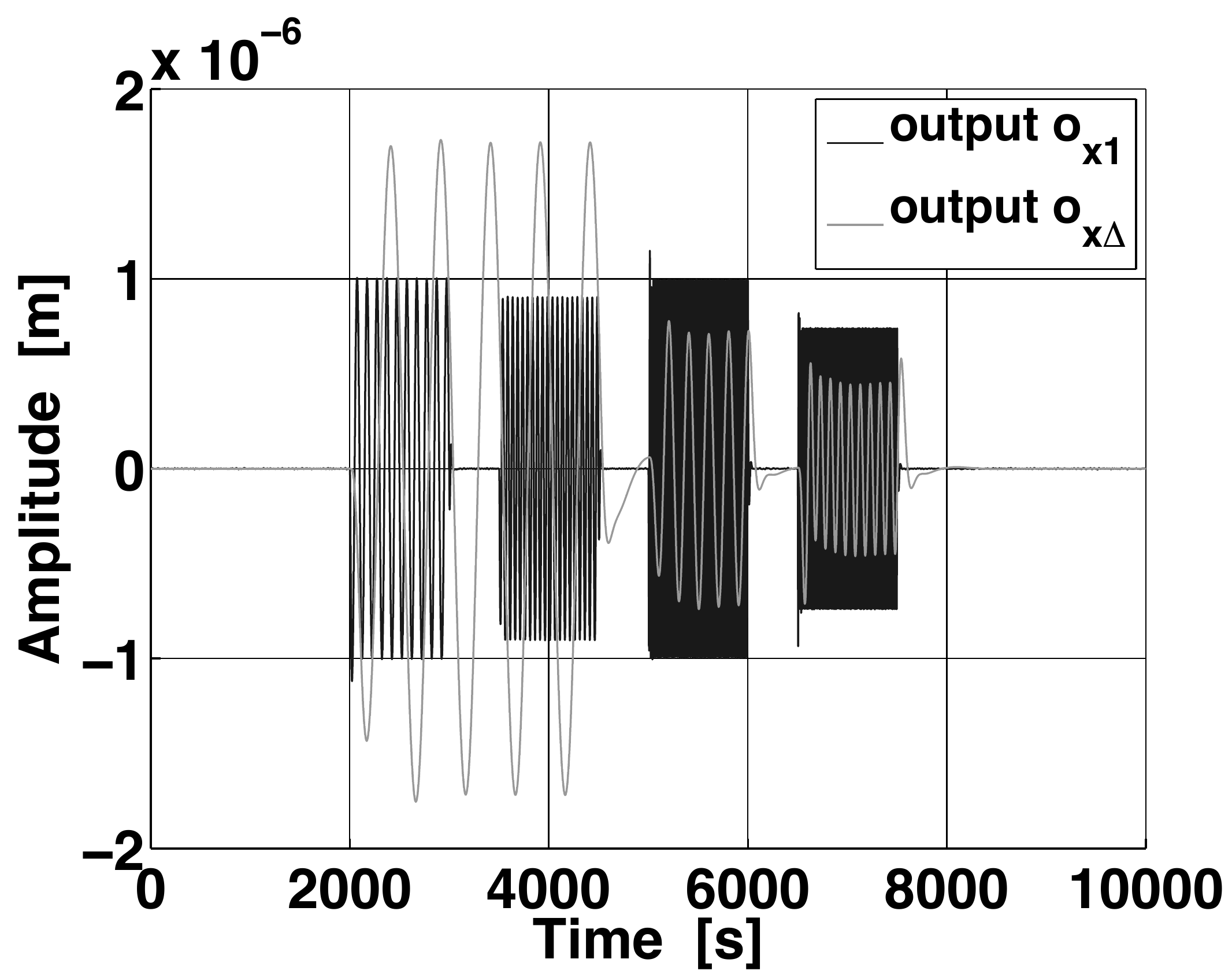}\\[0.2cm]
        \line(1,0){200}\; \textbf{Experiment 2}\; \line(1,0){200} \\[0.2cm]
        \includegraphics[width = 0.75\columnwidth]{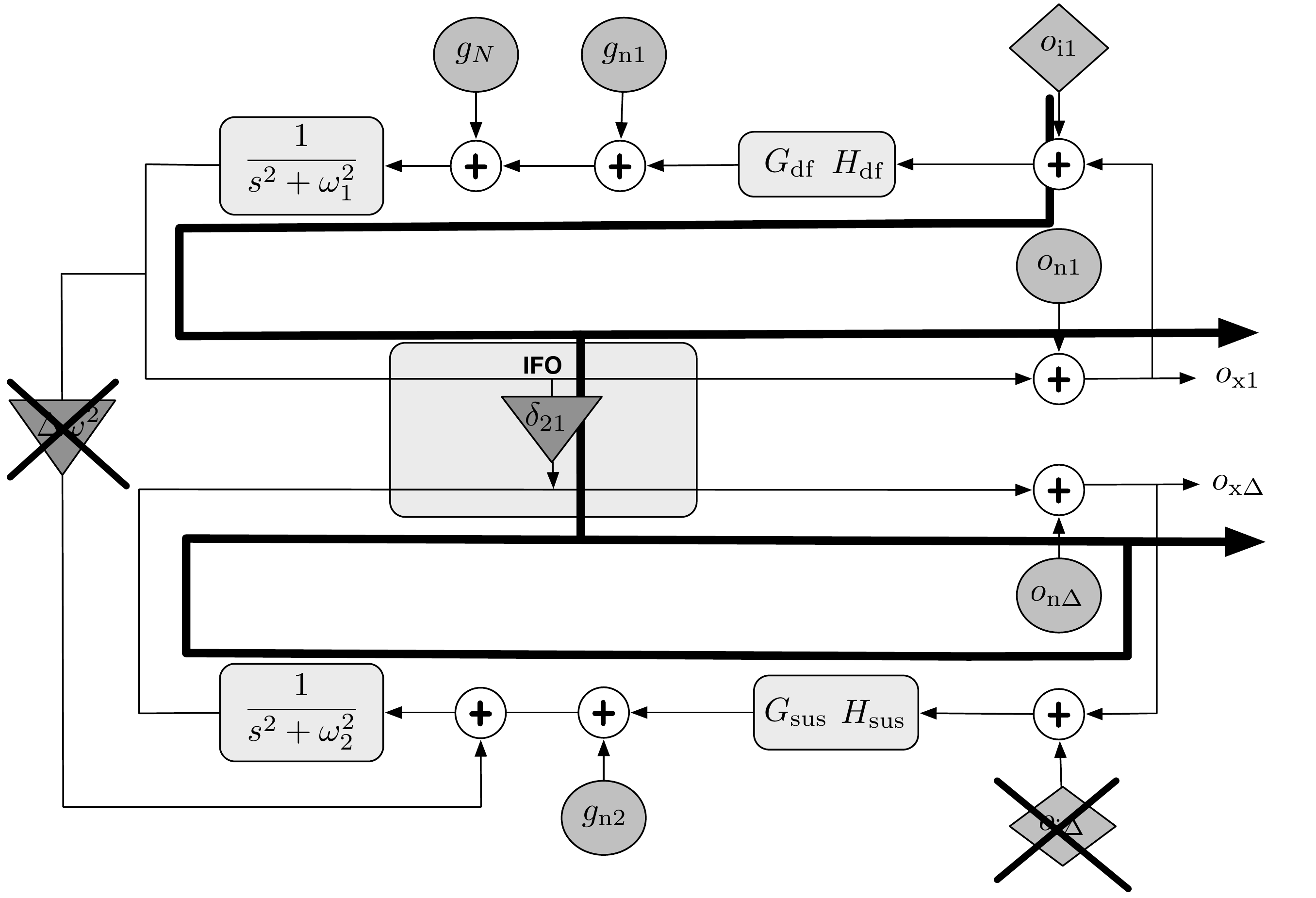}
        \includegraphics[width = 0.6\columnwidth]{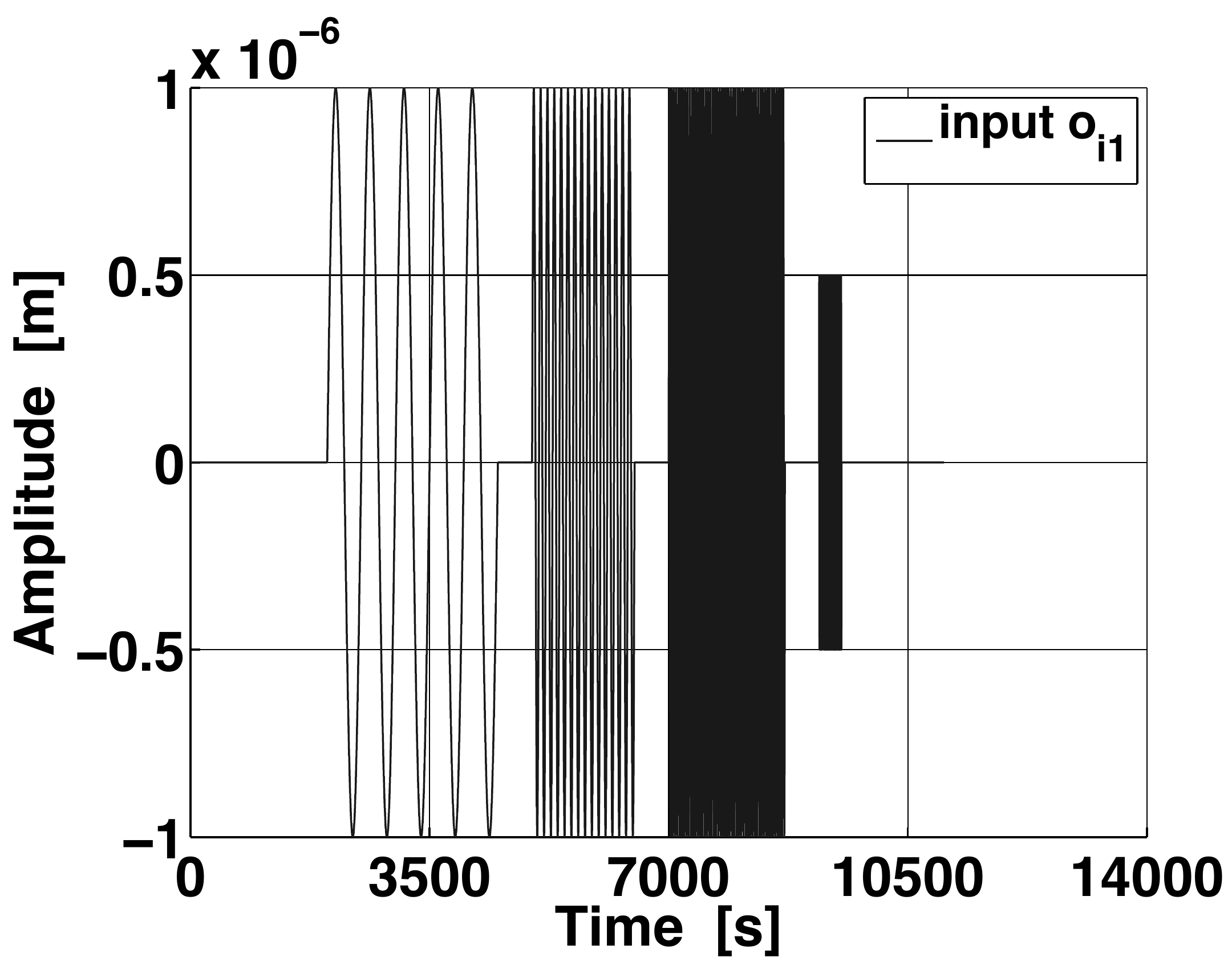}
        \includegraphics[width = 0.6\columnwidth]{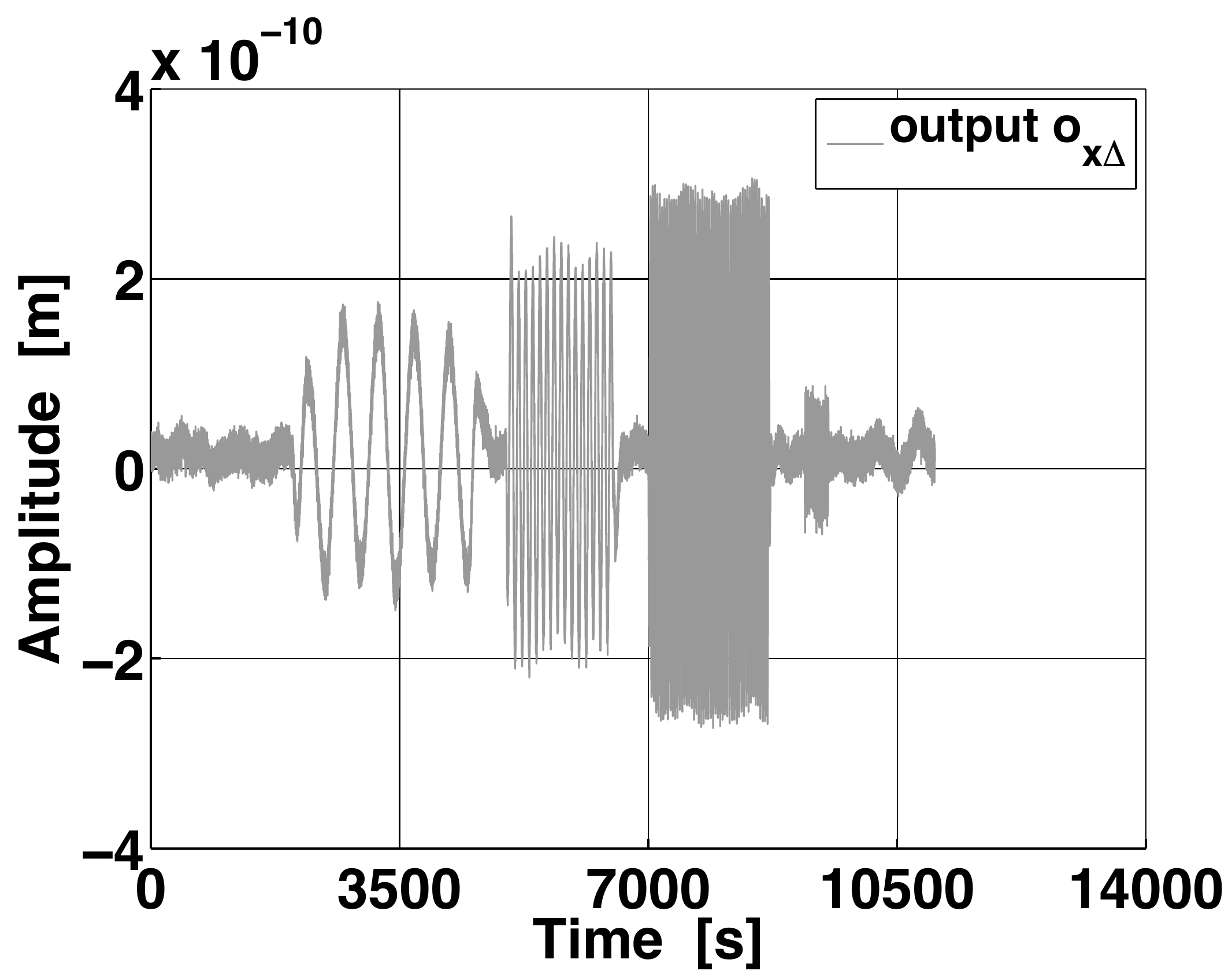}\\[0.2cm]
        \line(1,0){200}\; \textbf{Experiment 3}\; \line(1,0){200} \\[0.2cm]
        \includegraphics[width = 0.75\columnwidth]{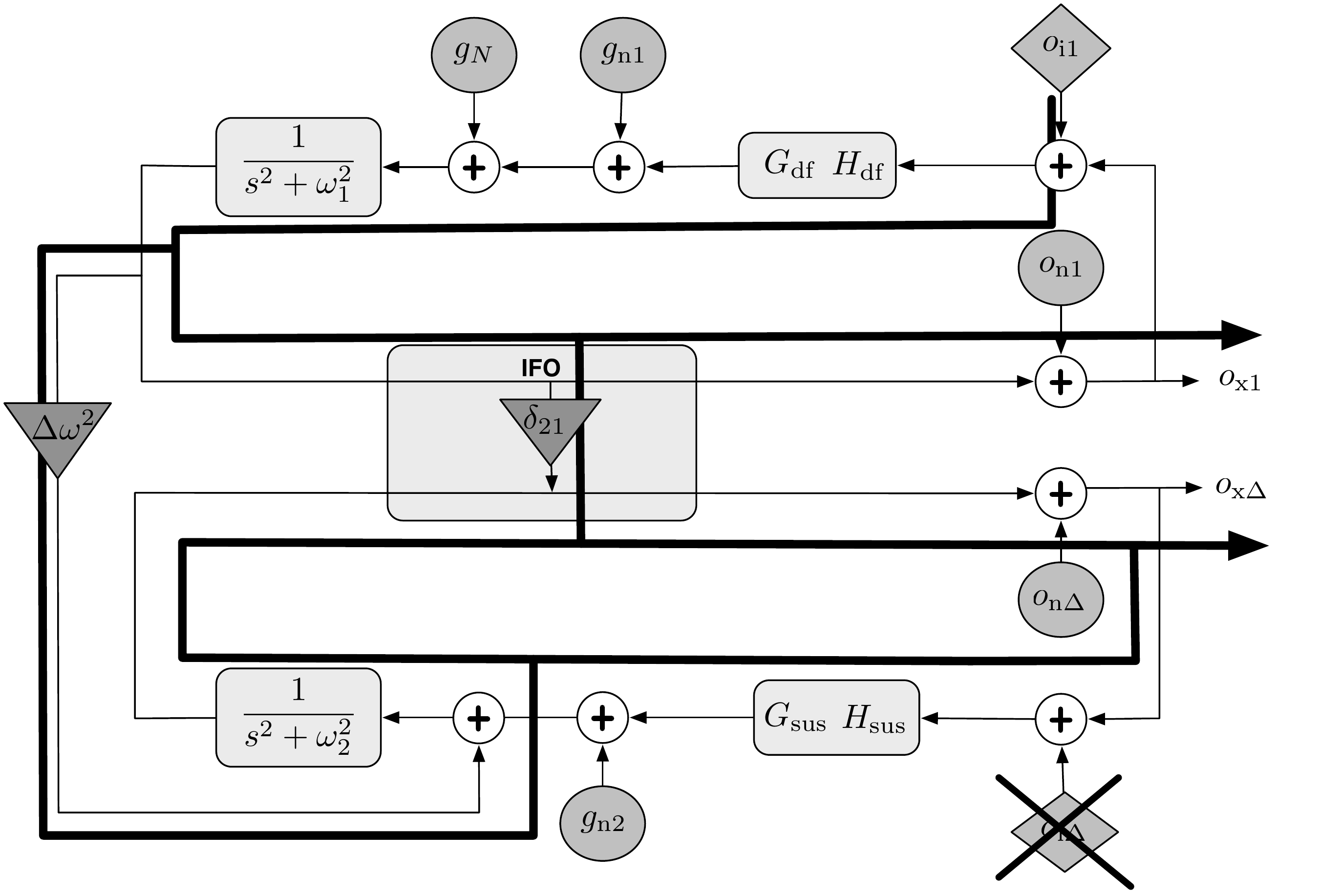}
        \includegraphics[width = 0.6\columnwidth]{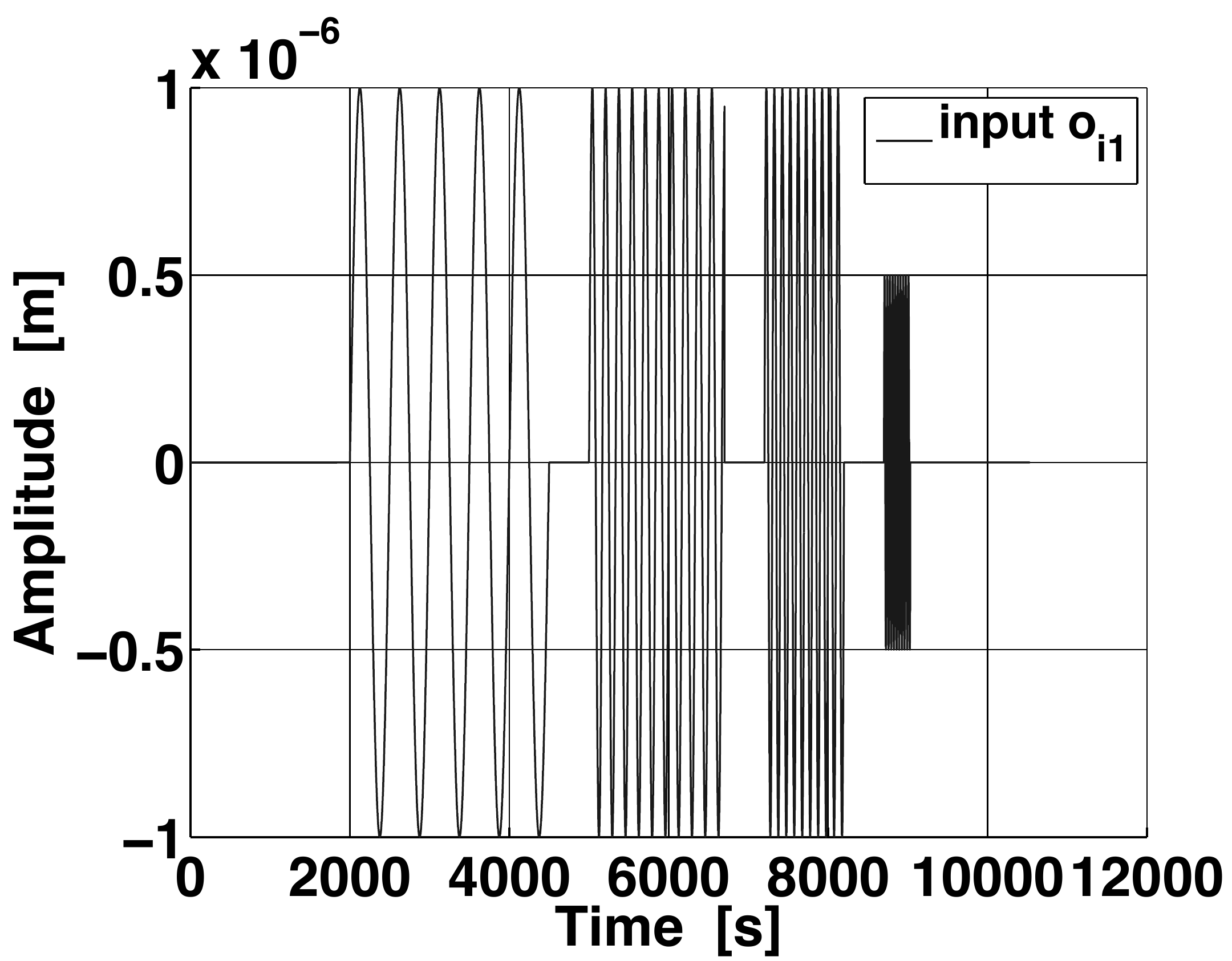}
        \includegraphics[width = 0.6\columnwidth]{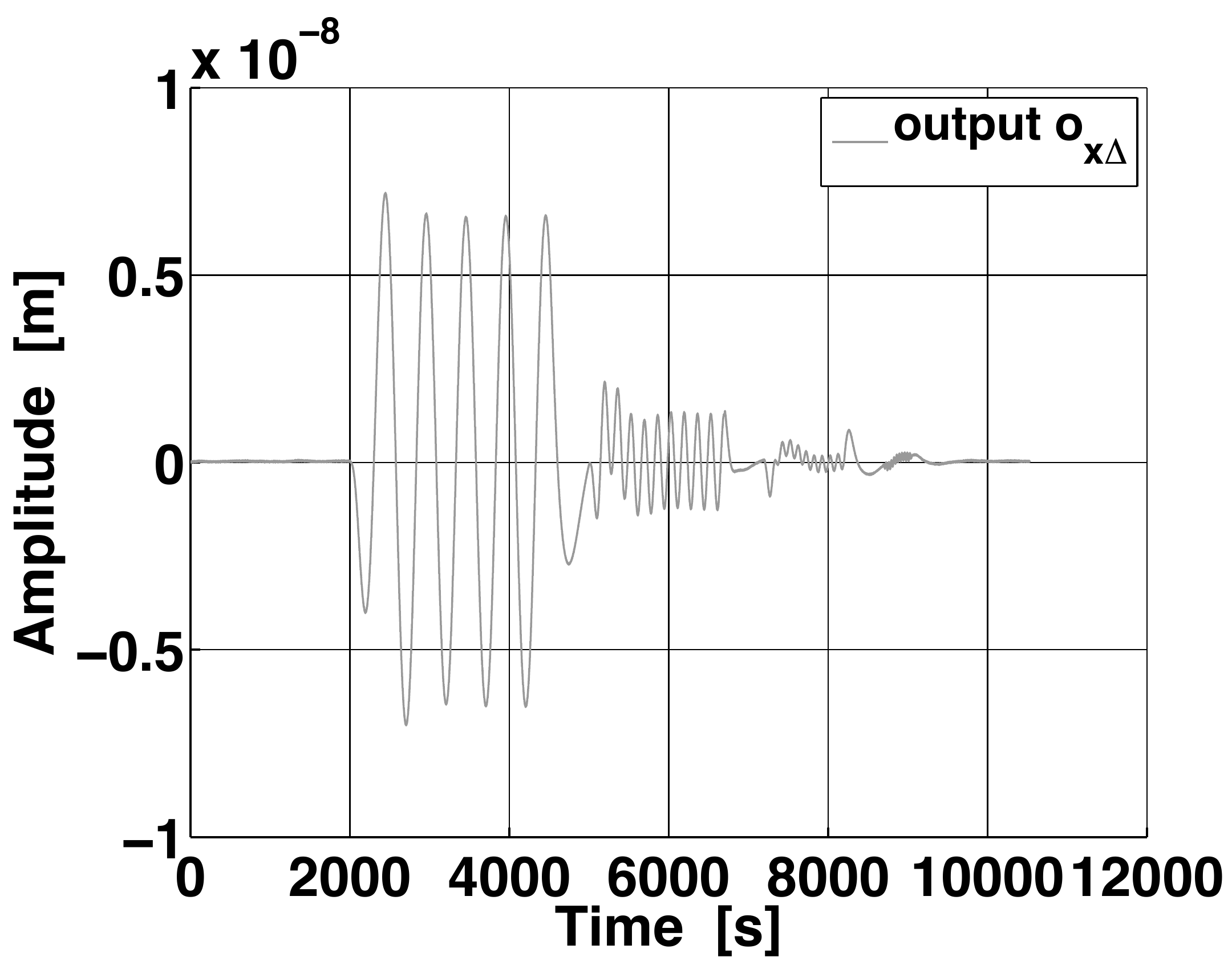}
      \caption{The three MDC2 experiments. From left to right: scheme of injected signal, input signal and output signal.
 From top to bottom: experiment 1, 2 and 3. Only $x_{12}$ output is shown for experiments 2 and 3, the response of 
 the first channel to the injected signal is similar to the one shown in experiment 1. \label{fig.signaltest1} }
  \end{center}
\end{figure*}

\begin{description}  
  \item[Experiment 1] Two signals are injected independently into the
    first and second channel. Each signal is a sequence of sinusoids
    with different amplitudes, frequencies and duration, all of them
    known to the data analysis team. This experiment is the richer in
    terms of frequencies injected to the system, and the one with best
    expected parameter estimates, as we show in the following section.
	
  \item[Experiment 2] A signal is injected in the first channel and
    both test mass stiffnesses are set to the same value,
    different than the value for the two other experiments. This
    configuration represents the \textsl{matched stiffness}
    configuration in the real LISA Pathfinder satellite. This state
    can be achieved by commanding an equal bias voltage on the
    electrodes of the inertial sensors at a level which dominates all
    other stiffness effects thus resulting in an equal coupling
    between the two test masses and the spacecraft. This scheme is
    particularly useful since it would ideally decouple any external
    force from the differential measurement. However, in our
    simplified model there is already a second cross-term, the
    interferometer cross-coupling, $\delta_{21}$, mixing both channels
    --- see Equation~(\ref{eq.xcoupl}). Being the only remaining
    cross-coupling in this experiment, this parameter should therefore
    be obtained with the greatest accuracy when analyzing this data
    set.

  \item[Experiment 3] The last experiment again applies only one
    signal to the first channel but without matching the stiffness for
    both test masses. This experiment tests the ability to recover the
    same parameters that we determine in Experiment 1, but by only
    injecting signals into the first channel.
\end{description}

The data set in MDC2 also included a run without any injected 
signal from where the instrument performance could be evaluated.
A typical noise realisation for this model is shown in 
Figure \ref{fig.noiseASD} whereas the three MDC2 experiments
are represented in Figure \ref{fig.signaltest1}, all of them 
generated using LTPDA methods. The concept
behind the data generation process is to translate the transfer
functions appearing in Equation~(\ref{eq.oo_all}) into digital
filters, and then use those filters to translate the input signal into
the measured output. Since the measured data is a combination of
signal and noise, the data generation procedure is consequently split
into two branches that are added at the end. The generation of the
signal part is straightforward since it only requires the filtering of
a deterministic signal. In contrast, the noise part requires the use of
digital filters to color white-noise and to do it in such a way that
the noise cross-correlation properties between the two channels are
correctly reproduced. A detailed description of this process can be
found in~\cite{Ferraioli10}.

% $Id: analysis.tex,v 1.28 2010/06/16 12:55:20 miquel Exp $

\section{Data analysis}

\subsection{Bayesian estimation}
 \label{sec.bayes}

We would now like to infer unknown parameters from the simulated data.
To this end we need to derive the \textsl{posterior probability
distribution} of the parameters, that is, the conditional probability
distribution of the parameters for the given data at hand.  The
posterior distribution expresses the information about the parameters
by assigning probabilities across parameter space, and by that allows
us to derive the most likely values and their uncertainties
\cite{Gelman,Gregory}.  The posterior distribution is given by
\textsl{Bayes' theorem}, and it depends on the data as well as any
other prior information~$I$:
\begin{equation} \label{eq.bayesTheorem}
 \prob(\Theta|D,I) 
 = \frac{\prob(\Theta|I) \times \prob(D|\Theta,I)}{\prob(D|I)}
 \propto \prob(\Theta|I) \times \prob(D|\Theta,I).
\end{equation}
The \textsl{prior probability distribution}~$\prob(\Theta|I)$
expresses information we may have about the parameter values (in
addition to the data~$D$), while the \textsl{likelihood
function}~$\prob(D|\Theta,I)$ describes the probabilistic relationship
between parameters and the (noisy) measurements. The
\textsl{evidence}~$\prob(D|I)$ is usually not of concern for parameter
estimation purposes and constitutes a normalizing constant here. In
this work we will assume uniform prior distributions for all
parameters, \ie, the prior density~$\prob(\Theta|I)$ is constant
across the allowed region as defined in Table~\ref{tbl.model}.

Given the simplified model in Equation~(\ref{eq.oo_all_simpl}) we
start by assuming that the noise term~$\vec{n}$ is Gaussian.  The
noise in each of the two output channels is characterized by the
(known) one-sided power spectral density functions $\speca(f)$ and
$\specb(f)$, respectively. In addition, the noise is assumed to be
correlated between the two outputs, which is expressed through the
cross spectral density~$\specab(f)$. Due to the colored noise it will
be convenient to express the likelihood function in terms of the
Fourier transformed data.  The likelihood function then is given by
%\begin{widetext}
\begin{eqnarray}
  \label{eq.likhd}
  p(D|\Theta,I) &=& \left[ (2\pi)^{N/2}\,\det \mathbf{\Sigma}\right]^{-1/2}\, \\
  & &\hspace{-0.8cm} \times \exp \left[ - \frac{1}{2} {\bigl(\vec{o}  - \mathbf{G_s}(\Theta)\,\vec{o}_i\bigr)}^{T}  \mathbf{\Sigma}^{-1} \bigl(\vec{o} - \mathbf{G_s}(\Theta)\,\vec{o}_i \bigr) \right], \nonumber
\end{eqnarray}
%\end{widetext}
so that (up to a multiplicative factor) the logarithmic likelihood is
proportional to the quadratic form
\begin{equation} \label{eq.loglik}
  \log\bigl(p(D|\Theta,I)\bigr) \,\propto \,
    -{\textstyle \frac{1}{2}} \,{\bigl(\vec{o}  -
\mathbf{G_s}(\Theta)\,\vec{o}_i \bigr)}^{T}  \mathbf{\Sigma}^{-1} \bigl(\vec{o} - \mathbf{G_s}(\Theta)\,\vec{o}_i \bigr),
\end{equation}
where $\mathbf{\Sigma}$ is the covariance matrix of the (Fourier
domain) noise term~$\vec{n}$.  The covariance matrix entries are then
defined by the spectral and cross-spectral density values
corresponding to the Fourier frequencies. Most of $\mathbf{\Sigma}$'s
entries are zero (since only the terms corresponding to the same
Fourier frequency are correlated) and the quadratic form may be
rearranged so that $\mathbf{\Sigma}$ is of a block-diagonal form and
the likelihood expression simplifies to a sum over the blocks of
correlated terms at each frequency bin:
\begin{equation} \label{eq.loglik2}
 \log\bigl(p(D|\Theta,I)\bigr) \;\propto \;
 {\textstyle -\frac{1}{2}} \sum_{j}\,
 \mathrm{Re} \bigl({\resi_j}^T \,\Sigma_j^{-1}\, \resi_j\bigr),
\end{equation}
where $j=0,\ldots,N/2$ is an index over the Fourier frequencies~$f_j$,
and $\resi_j$ and $\Sigma_j$ denote the two (complex-valued) residual
terms and corresponding covariance matrix at frequency~$f_j$:
\begin{eqnarray}
  \resi_j &=&  \left(\begin{array}{c}
                       \bigl[ \tosubxa  - \left( G_{11}(\Theta)\,\tosubia + G_{12}(\Theta)\,\tosubib \right) \bigr] (f_j)\\
                       \bigl[ \tosubxb  -  \left( G_{21}(\Theta)\,\tosubia + G_{22}(\Theta)\,\tosubib \right)  \bigr] (f_j)\\
                     \end{array}\right), \nonumber\\
  \Sigma_j &=& \frac{N}{4\Delta_t}
               \left(\begin{array}{cc}
                       \speca(f_j)   &  \specab(f_j)^\ast\\
                       \specab(f_j)  &  \specb(f_j)\\
                     \end{array}\right).
\end{eqnarray}

%---------------------------------------------------------------------

%\subsection{Optimal error estimate}
\subsection{Optimal parameter estimation errors\label{sec.crb}}

In order to get an idea of what kind of information the simulated
experiments will provide, we will use the \textsl{Fisher
information} formalism to estimate the measurement errors to be
expected from the different experimental settings. The Fisher
information and the corresponding Cram\'{e}r-Rao bound (CRB) provide
an estimate of the measurement uncertainties to be expected in the
limit of a large signal-to-noise ratio (SNR) \cite{Vallisneri08}.  For
an unbiased estimate of $\Theta$, the CRB can be expressed as
\begin{equation}
\mathrm{cov} (\Theta) \ge \textbf{J}^{-1}(\Theta),
\end{equation} 
where $\textbf{J}(\Theta)$ is the Fisher information matrix.
For our particular case it will shown to be useful to use the
Cram\'er-Rao bound expressed as~\cite{Zeira90},
\begin{widetext}
\begin{equation} 
[\textbf{J}(\Theta)]_{lm}   = \sum_{j,k} \left[ \frac{1}{2\,\pi}\, \int^{\hfill \infty}_{-\infty} \, d \omega \, \frac{1}{S_{jk}(\omega, \Theta)}  \frac{\partial o_{j}(\omega, \Theta)}{\partial \theta_l } \, \frac{\partial o_{k}(\omega,\Theta)}{\partial \theta_m} 
  +\frac{T}{4\,\pi}  \int^{\hfill \infty}_{-\infty} \, d \omega\, \frac{1}{S_{jk}^2(\omega , \Theta)} \frac{\partial S_{jk}(\omega , \Theta)}{\partial \theta_l} \frac{\partial S_{jk}(\omega , \Theta)}{\partial \theta_m} \right]
\label{eq.crb}
\end{equation}
\end{widetext}
where we sum over the two channels; $o_{x1}$ and $o_{x \Delta}$ 
being the two components of the nominal output, $S_{jk}(\omega, \mathbf{\Theta})$ 
the components of the cross-spectrum matrix
and $T$ the integration time. 
We are considering here the parametric 
dependence of the noise terms --- Equation~(\ref{eq.oo_all}).
Although we will drop it in the next step, we want to explicitly state
that term since it is usually not considered in the Fisher matrix
analysis among the gravitational wave community~\cite{Vallisneri08},
but it may turn out to be relevant in future analysis since the noise
model characterization is the final purpose of the LTP mock data
challenges. However, for this first application, and to avoid
cumbersome equations, we decided not to include those terms
considering that they will not introduce any relevant information in
the high SNR regime where we are working. Switching therefore to
Equation~(\ref{eq.oo_all_simpl}) and substituting into
Equation~(\ref{eq.crb}) leads to
\begin{eqnarray} 
  \label{eq.MDCcrb}
  && [\textbf{J}(\Theta)]_{lm} \\  
  &=&  \sum_{j,k} \frac{o_{i,j}\,o^*_{i,k}}{2\, \pi}
\int^{\hfill \infty}_{-\infty} \, d \omega \, \frac{1}{S_{jk}(\omega)}  \frac{\partial G_{jk}(\Theta)}{\partial \theta_l } \, \frac{\partial G_{jk}(\Theta)}{\partial \theta_m}, \nonumber
\end{eqnarray}
where now $o_{i,1}$ and $o_{i,\Delta}$ 
are the two components of the input
signal and $G_{jk}(\mathbf{\Theta})$ the components of the transfer function.
We will use Equation~(\ref{eq.MDCcrb}) in the following to
evaluate the CRB in each experiment. It is important to keep in mind
that the three experiments analyzed here contain different
configurations of the instrument, meaning that both the transfer
function elements and the signals
are changing in each experiment.
\begin{table}[h!]
\caption{Cr\'amer-Rao bound. Values between parenthesis expressed in relative parts per thousand ($\permil$) \label{tbl.crb}}
\begin{ruledtabular}
\begin{tabular}{c|ccc}
Parameter & Exp. 1 & Exp. 2 & Exp. 3 \\
\hline
 $\sigma_{G_{\rm df}}$ & ${2 \times 10^{-5}}\,(0.02)$       &  $5 \times 10^{-5}\,(0.06)$    &  $2 \times 10^{-4}\,(0.2)$ \\
 $\sigma_{G_{\rm sus}}$ &  ${3 \times 10^{-7}}\,(0.0002)$ &  $3 \times 10^{-3}\,(3)$        &  $3 \times 10^{-4}\,(0.3)$ \\
 $\sigma_{\omega_1}$  &  ${6 \times 10^{-10}}\,(0.5)$      & $3 \times 10^{-6}\,(1000)$   &   $9 \times 10^{-8}\,(80)$  \\
 $\sigma_{\omega_2}$ & ${3\times 10^{-10}}\,(0.1)$         &  $3 \times 10^{-6}\,(1000)$  &   $9 \times 10^{-8}\,(40)$  \\
 $\sigma_{\delta_{21}}$ & $6\times 10^{-8}\,(0.5)$           &    ${4\times 10^{-8}}\,(0.2)$   &  $1 \times 10^{-7}\,(0.9)$  \\
 $\sigma_{\Delta \omega}$ & ${5 \times 10^{-10}}\,(0.4)$    &  $6 \times 10^{-10}\,(-)$       &  $3 \times 10^{-10}\,(0.3)$ 
\end{tabular}
\end{ruledtabular}
\end{table}

Table~\ref{tbl.crb} summarizes the optimal error estimates that the
data analysis should return. 
The last column refers to the achievable
standard deviation in the difference between squared stiffnesses,
$\Delta \omega^2~=~\omega_2^2-\omega_1^2$.  This will be only
indirectly estimated by the analysis, but we added it to the table,
firstly, because the cross-coupling between both channels depends
directly on this difference, but also because the error in the
estimation of the stiffnesses difference depends on the non-diagonal
terms of the covariance matrix. This quantity adds then some
more information not contained in the other parameters, which are
extracted purely from the diagonal terms. The $\sigma_{\Delta \omega}$
error is computed as
\begin{equation}
\sigma^2_{\Delta \omega} = \sigma^2_{\omega_1}  + \sigma^2_{\omega_2} - 2
\sigma_{\omega_1,\omega_2},
\end{equation}
where $\sigma^2_{\omega_1}$ and $\sigma^2_{\omega_1}$ are the
variances of the stiffness squared of test mass 1 and test mass 2, and
$\sigma_{\omega_1,\omega_2}$ is the covariance term containing the
correlation between both stiffnesses. A remarkable result from this
analysis is that a single experiment injecting a signal in both
channels (experiment~1) is enough to determine all parameters with
high precision.  In fact, this experiment is preferable to the other
experiments which only inject signals in the $x_1$ channel. Only the
matched stiffness experiment (experiment~2) gives a slightly better
estimation of the interferometer cross-coupling. Precision
in this parameter is gained however at expenses of increasing
the uncertainty in the determination of the absolute value of 
the stiffnesses, reaching in this case 100\,\%. In principle, if we
take into account our simplified model, experiment~3 would be
redundant, not adding more information (apart from statistical
averaging) than what we get from experiments~1 and~2.

In order to give some more insight in what refers
the difference between experiments we provide in
Table \ref{tbl.corr} the correlation matrices as
computed with the previous formalism. These 
results complement the ones in Table \ref{tbl.crb}, 
since the diagonal terms of the latter
correspond to the values reported in the former.
Comparison between correlation matrices
show how experiment~1
is disentangling the different parameteres dependences 
more efficiently. In particular, it is the only experiment
which is able to differentiate the contribution of the two
stiffnesses. The reason for that being that it is the only 
experiment with a signal injected in the differential channel.
%A second important point to notice is that in all of them 
%the $\rm G_{\rm sus}$ parameter is strongly correlated 
%to both stiffnesses. This feature, intrinsic to the model, 
%is enhanced in the two last experiments. We will come 
%back to this when discussing the results of our analysis
%in section~\ref{sec.discussion}. 
%
\begin{table}[h!]
\caption{Correlation matrices for MDC2 experiments \label{tbl.corr}}
\begin{ruledtabular}
\begin{tabular}{cccccc}
& $\rm G_{df}$ & $\rm G_{sus}$  & $\rm \omega^2_{1}$   & $\rm \omega^2_{2}$  & $\rm \delta_{21}$ \\
 \cline{2-6}
&  \multicolumn{5}{c}{Experiment 1}\\
 \cline{2-6}
$\rm G_{df}$             	&    1   	&   0.0003	&    -0.1	&	  -0.001	&        -0.2   \\
$\rm G_{sus}$   			&  0.0003 	&          1      &  -0.3     &   -0.5      	& -0.001		\\
$\rm \omega^2_{1}$    &   -0.1      &  -0.3         &      1      &   0.5         	& 0.5		\\
$\rm \omega^2_{2}$    &     -0.001   &    -0.5       &  0.5       &      1       	& 0.005		\\
$\rm \delta_{21}$         &   -0.2     & -0.001      &   0.5      &  0.005       &              1  \\
 \cline{2-6}
&  \multicolumn{5}{c}{Experiment 2}\\
 \cline{2-6}
 $\rm G_{df}$             	&            1     &    0.4       	&  -0.6						&    -0.6				&         0.2  \\
$\rm G_{sus}$   			&         0.4     &     1        	& -0.7      					& -0.7        			&	0.3		\\
$\rm \omega^2_{1}$    &        -0.6    &    -0.7        & 1         					&$\approx 1$       &	-0.4			\\
$\rm \omega^2_{2}$    &         -0.6    &   -0.7         & $\approx 1$           &1        				&		-0.4		\\
$\rm \delta_{21}$         &          0.2    &    0.3         & -0.4      				 & -0.4       			&		1		\\
 \cline{2-6}
 & \multicolumn{5}{c}{Experiment 3}\\
 \cline{2-6}
 $\rm G_{df}$             	&           1		&    0.03    	&   -0.02	&  -0.02	&         0.04 \\
$\rm G_{sus}$   			&        0.03    &       1         &-0.8       & -0.8      &   0.3        \\
$\rm \omega^2_{1}$    &       -0.02    &    -0.8       &   1         &$\approx 1$        	& -0.09      \\
$\rm \omega^2_{2}$    &       -0.02    &    -0.8       &  $\approx 1$          &  1       	& -0.09     \\
$\rm \delta_{21}$         &        0.04     &   0.3          & -0.09    &  -0.09   &       1     \\
\end{tabular}
\end{ruledtabular}
\end{table}

%---------------------------------------------------------------------

\subsection{Combining the results of experiments}
  \label{sec.combine}

\subsubsection{The information propagation problem}

As opposed to the usual application of Bayesian parameter estimation
in LISA, where a single set of data is used to determine the
parameters of a multiplicity of systems, \ie, astrophysical sources,
in our case we use different sets of data (experiments) to
characterize a unique system, the LTP experiment. Thus, once we have
obtained the parameter estimates for each experiment we still need to
go further to achieve our final goal. Since each experiment can be
adding valuable, but partial, information about the instrument, we
need to find a scheme that allows us to include all the information in
a final set of parameters.

The efficient combination of results is also an important problem to
solve in terms of mission operations. It should be noted that the LISA
Pathfinder mission will be a space laboratory with approximately 100 channels
being sampled and more than 50 parameters defining its performance.
It will therefore be crucial to combine the results from one
experiment with the ones following. For instance, we may be interested in
using the determination of the stiffness to calibrate the thrusters in
a forthcoming experiment. Given the limited mission time and the high
numbers of experiments to be performed, the need for a clear
combination scheme is evident. We explore in the following how to take
advantage of the posterior distribution to that end.

%---------------------------------------------------------------------

\subsubsection{The general case}

\paragraph{Identical parameter sets}

First consider the case where the parameter sets are identical for the
data sets to be combined (as e.g.\ in Experiments 1 and 3 above).  Suppose
we have a parameter vector~$\Theta$ and two data sets $D_1$ and $D_2$.
Similar to the general case in Equation~(\ref{eq.bayesTheorem}), the
posterior distribution $\prob(\Theta|D_1,D_2,I)$ is then given by
%\begin{eqnarray}\label{eqn.bayes.combined}
%  \prob(\Theta|D_1,D_2,I)&\propto &
%  \overbrace{\phantom{\prob(\Theta|I)}}^{\mbox{prior}}\phantom{\times}\overbrace{\phantom{\prob(D_1|\Theta,I)\times\prob(D_2|\Theta,I)}}^{\mbox{likelihood}}\nonumber\\[-14.5pt]
%  &&\underbrace{\prob(\Theta|I)\times\prob(D_1|\Theta,I)}_{\mbox{prior}}\times\underbrace{\prob(D_2|\Theta,I)}_{\mbox{likelihood}},
%\end{eqnarray}
\begin{eqnarray}\label{eqn.bayes.combined}
  \prob(\Theta|D_1,D_2,I)&\propto &
  \overbrace{\phantom{\prob(\Theta|I)}}^{\mbox{prior}}\phantom{\!\times\!}\overbrace{\phantom{\prob(D_1|\Theta,I)\!\times\!\prob(D_2|\Theta,I)}}^{\mbox{likelihood}}\qquad\nonumber\\[-14.5pt]
  &&\underbrace{\prob(\Theta|I)\!\times\!\prob(D_1|\Theta,I)}_{\mbox{prior}}\!\times\!\underbrace{\prob(D_2|\Theta,I)}_{\mbox{likelihood}},\qquad
\end{eqnarray}
where the same expression may be motivated by either taking the
likelihood to be the product of the individual experiments'
likelihoods or by analyzing the experiments one after the other and
using the posterior from the first experiment as the prior for the
second experiment (\ref{eqn.bayes.combined}).

\paragraph{Differing parameter sets}\label{sec.combine.different}

In order to deal with differing parameter sets that only partially
overlap, one needs to consider the union of all the unknowns as the
set of parameters. Combining data from different experiments then
works exactly as in Equation~(\ref{eqn.bayes.combined}), only that the
parameter vector~$\Theta$ is now the extended parameter set.  The
likelihood functions are exactly the same as in the
individual-experiment case, with the only difference that, as functions of the
extended parameter set, they do not depend on some of the parameters.

Consider the case where two data sets $D_1$ and $D_2$ 
depend on parameter~$\vartheta_1$, while the parameters
$\vartheta_2$ and $\vartheta_3$ are specific for $D_1$ and $D_2$,
respectively. Assuming the error terms for both experiments to be
independent, the joint likelihood function then is the product
%\begin{widetext}
\begin{eqnarray} \label{eqn.updated.posterior1}
 && \prob(D_1, D_2 | \vartheta_1, \vartheta_2, \vartheta_3, I)\nonumber\\
&=& \prob(D_1| \vartheta_1, \vartheta_2, \vartheta_3, I) \times
    \prob(D_2| \vartheta_1, \vartheta_2, \vartheta_3, I), \nonumber\\
&=& \prob(D_1| \vartheta_1, \vartheta_2, I) \times
    \prob(D_2| \vartheta_1, \vartheta_3, I).
\end{eqnarray}
%\end{widetext}
In order to simplify things, in the following we will introduce the
assumption that the conditional prior
$\prob(\vartheta_2|\vartheta_1,\vartheta_3,I)$ is independent of
$\vartheta_3$, \ie,
\begin{equation} \label{eqn.updated.posterior2.condition}
  \prob(\vartheta_2|\vartheta_1,\vartheta_3,I)\;=\;\prob(\vartheta_2|\vartheta_1,I)
\end{equation}
%(since $\vartheta_2$ and $\vartheta_3$ were the parameters which did
%not jointly affect both experiments, this may be easily satisfied).
(since $\vartheta_2$ and $\vartheta_3$ were the parameters which did
not jointly affect both experiments, this may be easily satisfied, for
example if
$\prob(\vartheta_1,\vartheta_2,\vartheta_3|I)=\prob(\vartheta_1|I)\times\prob(\vartheta_2|I)\times\prob(\vartheta_3|I)$).
When considering additional data~$D_2$, the change in the (marginal)
posterior distribution of the two parameters $\vartheta_1$ and
$\vartheta_2$ then is given by
\begin{equation}
\prob(\vartheta_1,\vartheta_2|D_1,D_2,I)
%  &=& \int \prob(\vartheta_1,\vartheta_2,\vartheta_3|D_1,D_2,I)\,\diff \vartheta_3 \nonumber\\
%  &=& \int \frac{\prob(D_1|\vartheta_1,\vartheta_2,I) \, \prob(D_2|\vartheta_1,\vartheta_3,I) \, \prob(\vartheta_1,\vartheta_2,\vartheta_3|I)}{\prob(D_1,D_2,I)}\,\diff \vartheta_3 \nonumber\\
%  &=& \frac{\prob(D_1|\vartheta_1,\vartheta_2,I) \, \prob(\vartheta_1,\vartheta_2,I)}{\prob(D_1,I)} \times \int \frac{\prob(D_2|\vartheta_1,\vartheta_3,I)\,\prob(\vartheta_3|\vartheta_1,\vartheta_2,I)}{\prob(D_2,I)}\,\diff \vartheta_3\nonumber\\
%  &=& \prob(\vartheta_1,\vartheta_2|D_1,I) \times \int \frac{\prob(D_2|\vartheta_1,\vartheta_3,I)\,\prob(\vartheta_3|\vartheta_1,\vartheta_2,I)}{\prob(D_2,I)}\,\diff \vartheta_3\nonumber\\
%  &=& \prob(\vartheta_1,\vartheta_2|D_1,I) \times \frac{1}{\prob(\vartheta_1|I)}\int \frac{\prob(D_2|\vartheta_1,\vartheta_3,I)\,\prob(\vartheta_1,\vartheta_3|I)}{\prob(D_2|I)}\,\frac{\prob(\vartheta_2|\vartheta_1,\vartheta_3,I)}{\prob(\vartheta_2|\vartheta_1,I)}\,\diff \vartheta_3 \nonumber\\
  =  \prob(\vartheta_1,\vartheta_2|D_1,I) \times \frac{\prob(\vartheta_1|D_2,I)}{\prob(\vartheta_1|I)},
    \label{eqn.updated.posterior2}
\end{equation}
so that in order to ``update'' the posterior distribution of
$\vartheta_1$ and $\vartheta_2$ using the data $D_2$ that depends on
the additional parameter $\vartheta_3$, we only need to consider the
\textsl{marginal} prior and posterior distributions of the common
parameter~$\vartheta_1$, $\prob(\vartheta_1|I)$ and
$\prob(\vartheta_1|D_2,I)$. We can see that when updating the
posterior by another posterior (\ref{eqn.updated.posterior2}), the
(marginal) prior needs to be cancelled out, otherwise it would enter
twice into the resulting posterior.  Since by combining the posteriors
we will only learn about the common parameter~$\vartheta_1$ here, it
will be easier to also integrate out~$\vartheta_2$ and only consider
the (marginal) distributions involving~$\vartheta_1$, which then leads
to
\begin{equation}
  \prob(\vartheta_1|D_1,D_2,I)
  =  \prob(\vartheta_1|D_1,I) \times \frac{\prob(\vartheta_1|D_2,I)}{\prob(\vartheta_1|I)}.
    \label{eqn.updated.posterior3}
\end{equation}
The higher-dimensional case works
completely analogously, just by considering the parameters
$\vartheta_1$, $\vartheta_2$, $\vartheta_3$ to be sub-vectors.

%---------------------------------------------------------------------

\subsubsection{The Gaussian approximation}

As we will see below, the derived posterior distributions often turn
out to be well approximated by a multivariate Gaussian distribution
with mean~$\mathbf{\mu}$ and covariance matrix~$\mathbf{\Sigma}$:
\begin{eqnarray}
  \label{eq.mvgauss}
  &   & p(\mathbf{x}|D) \;\approx\; p(\mathbf{x; \mu, \Sigma}) \nonumber \\
  & = & \frac{1}{(2\,\pi)^N \, \vert \Sigma \vert ^{1/2}} \exp \Bigl \lbrace -\frac{1}{2} (\mathbf{x-\mu})^T \mathbf{\Sigma^{-1}} (\mathbf{x-\mu}) \Bigr \rbrace.
\end{eqnarray}
If the posterior distributions are expressed as Gaussians, it is
particularly easy to \textsl{analytically} propagate prior and
posterior information as described in the previous subsection; in the
following we will therefore apply these results to the Gaussian case.
As a further simplification, we will also assume all prior
distributions to be uniform.
%\begin{figure*}[t!]
%\begin{center}
%        \includegraphics[scale = 0.28]{./images/chain1_exp1_bw}
%        \includegraphics[scale = 0.28]{./images/chain5_exp1_bw}
%        \\[0.5cm]
%        \includegraphics[scale = 0.28]{./images/chain3_exp1_bw}
%        \includegraphics[scale = 0.28]{./images/chain4_exp1_bw}
%      \caption{MCMC for experiment 1. The chains are moving in the first 2000 
%      samples (search phase) from the starting values to the true ones. 
%      Since these initial values are the ones obtained by the simplex algorithm, 
%      the example shows how the MCMC is able to correct from the
%      secondary maxima originally obtained by the simplex.\label{fig.Exp1chain}}
% \end{center}
%\end{figure*}
%\begin{figure*}[t!]
%---------------------------------------------------------------------

\paragraph{Identical parameter sets} 

In order to combine the results coming from two experiments $D_1$ and
$D_2$, we will need to combine their two posterior distributions as in
Equation~(\ref{eqn.bayes.combined}). The results from experiments~$D_1$
and~$D_2$ will be summarized by parameters' posterior means and covariances 
$\lbrace \mathbf{\mu_1}, \mathbf{\Sigma_1} \rbrace $ and $\lbrace
\mathbf{\mu_2},\mathbf{\Sigma_2} \rbrace $, respectively.
Assuming uniform priors, we can now combine both as
\begin{eqnarray}
  p(\mathbf{x} | D_1,D_2)  
  & = & p(\mathbf{x} | D_1) \times p(\mathbf{x} | D_2) \nonumber\\
  & = & p(\mathbf{x; \mu_1,\Sigma_1}) \times p(\mathbf{x; \mu_2,\Sigma_2}) \nonumber\\
  & = & p(\mathbf{x; \mu_c,\Sigma_c}),
\end{eqnarray}
i.e., the product of posterior densities again is Gaussian with
mean~$\mathbf{\mu_c}$ and covariance~$\mathbf{\Sigma_c}$.  The
parameters of the combined posterior may then be derived using the
following relationship
\begin{eqnarray}
& &\rm \bf (x-u)^T U^{-1} (x-u) + (x-v)^T V^{-1} (x-v), \nonumber \\
&=&\rm \bf (x-w)^T W^{-1} (x-w), 
\end{eqnarray}
where
\begin{equation}
\rm \bf  w = W^{-1} [Uu + Vv],\quad W = U + V, 
\end{equation}
so that the new mean and covariance turn out as
\begin{eqnarray}
\bf \Sigma_{c}^{-1} & = & \bf \Sigma_{1}^{-1}  + \Sigma_{2}^{-1} \\
\bf \mu_{c} & = & \bf \Sigma_{c} \left[  \Sigma_{1}^{-1} \mu_1 +  
\Sigma_{2}^{-1} \mu_2\right]
\end{eqnarray}
\cite{Gelman}.  The same argument is easily extended to an arbitrary
number~$N$ of experiments as
\begin{eqnarray}
\bf \Sigma_{N}^{-1} & = & \sum^{N}_{i=1} \mathbf{\Sigma}_{i}^{-1}  \\
\bf \mu_{N} & = & \mathbf{\Sigma_{N}} \, \sum^{N}_{i=1} 
\mathbf{\Sigma}_{i}^{-1}  \mathbf{\mu}_i.
\label{eq.newjoint}
\end{eqnarray}

%---------------------------------------------------------------------

\paragraph{Differing parameter sets} 

Now suppose we have results of two experiments in which the parameter
sets were not quite identical, as in the previous
Section~\ref{sec.combine.different}.  One may now either directly
derive estimates of the \textsl{marginal} distribution (i.e., their
means and covariances) and use those to combine the marginal
posteriors as in Equation~(\ref{eqn.updated.posterior3}) and in the
previous section.  Otherwise, if given only the joint distributions
(means and covariances) of the differing (but intersecting) parameter
sets, these may also be marginalized analytically. For a Gaussian
distribution the marginal distribution of a subset of the variables is
simply given by the corresponding subset of mean and covariance
parameters, i.e., by dropping the rows and columns corresponding to
the variables that are integrated out.
\begin{table*}[t!]
\caption{Estimated parameters for independent experiments. \label{tbl.Exps}}
\begin{ruledtabular}
\begin{tabular}{lcccc}
Param. & Value~$\theta$  & Estimated~$\hat{\theta}\pm\sigma$ & $\vert \theta-\hat{\theta}\vert/\sigma$ & $\sigma/\sigma_{\rm CRB}$\\
\hline
 & \multicolumn{4}{c}{Experiment 1}\\
\cline{2-5}
$\rm G_{df}$ 				& 0.8  								&  $ 0.800\,02 \pm 0.000\,02$  						& 1.0 & 1.0 \\
$\rm G_{sus}$  				& 1.15  							&  $ 1.150\,000\,1 \pm 0.000\,000\,3 $   	    & 0.4 & 0.9 \\
$\rm \omega^2_{1}$   	& $-1.1 \times 10^{-6}$  	&  $(-1.099\,1 \pm 0.000\,5) \times 10^{-6}$  & 1.7 & 1.0\\
$\rm \omega^2_{2}$   	& $-2.2 \times 10^{-6}$  	&  $(-2.200\,1 \pm 0.000\,3) \times 10^{-6}$ 	& 0.3 & 1.0\\
$\rm \delta_{21}$  		& $1.35 \times 10^{-4}$	&$ (1.350\,2 \pm 0.000\,6)  \times 10^{-4}$ 	& 0.3 & 1.0 \\
$\rm \Delta \omega^2$ & $-1.1 \times 10^{-6}$ & $(-1.101\,0 \pm 0.000\,5) \times 10^{-6} $  	& 2.1 & 1.0\\
 & \multicolumn{4}{c}{Experiment 2}\\
 \cline{2-5}
$\rm G_{df}$ 				& 0.8 									& $ 0.800\,11 \pm 0.000\,05 $ 						& 2.2 & 1.0 \\
$\rm G_{sus}$  				& 1.15  								&  $ 1.147 \pm 0.004 $ 									& 0.8 & 1.0 \\
$\rm \omega^2_{1}$   	&   $-2.4  \times 10^{-6}$		&      $(-5 \pm 3)\times 10^{-6}$						& 0.8 &  1.0 \\
$\rm \omega^2_{2}$   	&  $-2.4  \times 10^{-6}$  	&      $(-5 \pm 3)\times 10^{-6}$ 						& 0.8 &  1.0 \\
$\rm \delta_{21}$  		& $1.35 \times 10^{-4}$ 		&$ (1.349\,7 \pm 0.000\,3)  \times 10^{-4}$ 	& 1.0 & 0.9 \\
$\rm \Delta \omega^2$ & $0$ 									& $(-3 \pm 6) \times 10^{-10}$						& 0.5 & 1.1\\
 & \multicolumn{4}{c}{Experiment 3}\\
 \cline{2-5}
$\rm G_{df}$ 				& 0.8 								& $ 0.799\,8 \pm 0.000\,2 $ 								& 1.2 & 1.1\\
$\rm G_{sus}$  				& 1.15  							&  $ 1.150\,3 \pm 0.000\,3 $ 								& 0.8 & 1.0\\
$\rm \omega^2_{1}$   	&  $-1.1  \times 10^{-6}$	&  $(-1.25 \pm 0.09) \times 10^{-6}$ 					& 1.7 & 1.0 \\
$\rm \omega^2_{2}$   	&  $-2.2  \times 10^{-6}$ &  $(-2.35 \pm 0.09) \times 10^{-6}$ 					& 1.7 & 1.0 \\
$\rm \delta_{21}$  		&  $1.35 \times 10^{-4}$ 	&$ (1.350 \pm 0.001)  \times 10^{-4}$ 				& 0.3 & 1.0 \\
$\rm \Delta \omega^2$ & $-1.1  \times 10^{-6}$ & $(-1.0999 \pm 0.0003) \times 10^{-6}$ 			& 0.2 & 1.0\\
\end{tabular}
\end{ruledtabular}
\end{table*}
%---------------------------------------------------------------------
\subsection{Implementation}

Our implementation follows a four-step procedure to analyze each
experiment, all of them implemented as \mbox{LTPDA} methods. The first
step is to Fourier transform the data.  The noise's power spectral
density is estimated using the Welch method~\cite{Welch67} and
applying a Blackman-Harris window.  We can then compute the
log-likelihood (\ref{eq.likhd}) and therefore find the maximum of the
posterior density function using a (Nelder-Mead) simplex search
algorithm \cite{Himmelblau}.  Since with our strong signal injections
the likelihood surface apparently does not tend to exhibit many
secondary maxima, this step is usually sufficient to determine the
parameters to good accuracy and it is also more efficient than waiting
for the Metropolis sampler to converge.  However, if the likelihood
surface shows secondary maxima, this method may lead to an erroneous
result. Next, the posterior covariance among parameters
according to input signals, noise and the relevant transfer functions,
is estimated by numerically evaluating the Fisher information matrix
at the maximum determined in the previous optimization step. And
finally, we can integrate the posterior using a Markov Chain Monte
Carlo (MCMC) approach. We use a Metropolis algorithm
\cite{Gelman,GilksRichardson} that will generate random samples from
the parameters' (5-dimensional) posterior distribution. Generation of
these samples is relatively easy based only on the expression of the
(unnormalized) posterior density function ((\ref{eq.bayesTheorem})
or~(\ref{eq.loglik})).

In order to enhance convergence of the MCMC sampler, we apply
\textsl{tempering} to the posterior density function, which is
supposed to make it more tractable and keep the algorithm from getting
stuck in local optima. In the MCMC context, tempering is commonly
implemented by applying an exponent to the probability density to be
sampled from, i.e., instead of using the posterior~ $p(\theta|D,I)$,
the tempered posterior~$p(\theta|D,I)^\frac{1}{T}$ is considered,
where $T \geq 1$ is the ``temperature''
\cite{GilksRichardson,Roever2007}.  The $\frac{1}{T}$~exponent
smoothens the targeted density function, which generally allows the
sampler to move more quickly and widely through parameter space and to
traverse between local modes more easily. The following expression
describes the temperature profile used in our
implementation~\cite{Cornish07},
\begin{equation}
T = \left\lbrace
  \begin{array}{cc}
  10^{\xi \left (1 - \frac{\rm T_h}{\rm T_c} \right)}& 1 \le i \le {\rm T_h} \\
  10^{\xi \left( 1 - \frac{i}{\rm T_c} \right)}  & {\rm T_h} \le i \le {\rm T_c} \\
   1 &  i \ge {\rm T_c},
  \end{array} \right.
\end{equation}
with $i$ indexing the samples of the Metropolis chain.  We initially
applied a constant temperature (with $\xi=3$) for the first
1000~iterations ($\rm T_h = 1000$), which was then exponentially
annealed down in the following 1000~iterations ($\rm T_c = 2000$),
after which the algorithm was properly generating samples from
the actual posterior distribution. To reduce the time required during
the search phase we occasionally rescale the covariance matrix of the
proposal distribution to explore a wider region of the parameter
space. Also, as proposed in~\cite{Gelman,Cornish07}, we correct the
standard deviation of the proposal distribution with a factor of
$\rm d^{-1/2}$, where d is the parameter space dimension.

%---------------------------------------------------------------------
\subsection{Results and discussion \label{sec.discussion}}

\begin{figure*}[p]
\begin{center}$
\begin{array}{ccc}
  \textbf{Experiment\;1} & \textbf{Experiment\;2} & \textbf{Experiment\;3}  \\[0.2cm]
 \includegraphics[scale = 0.19]{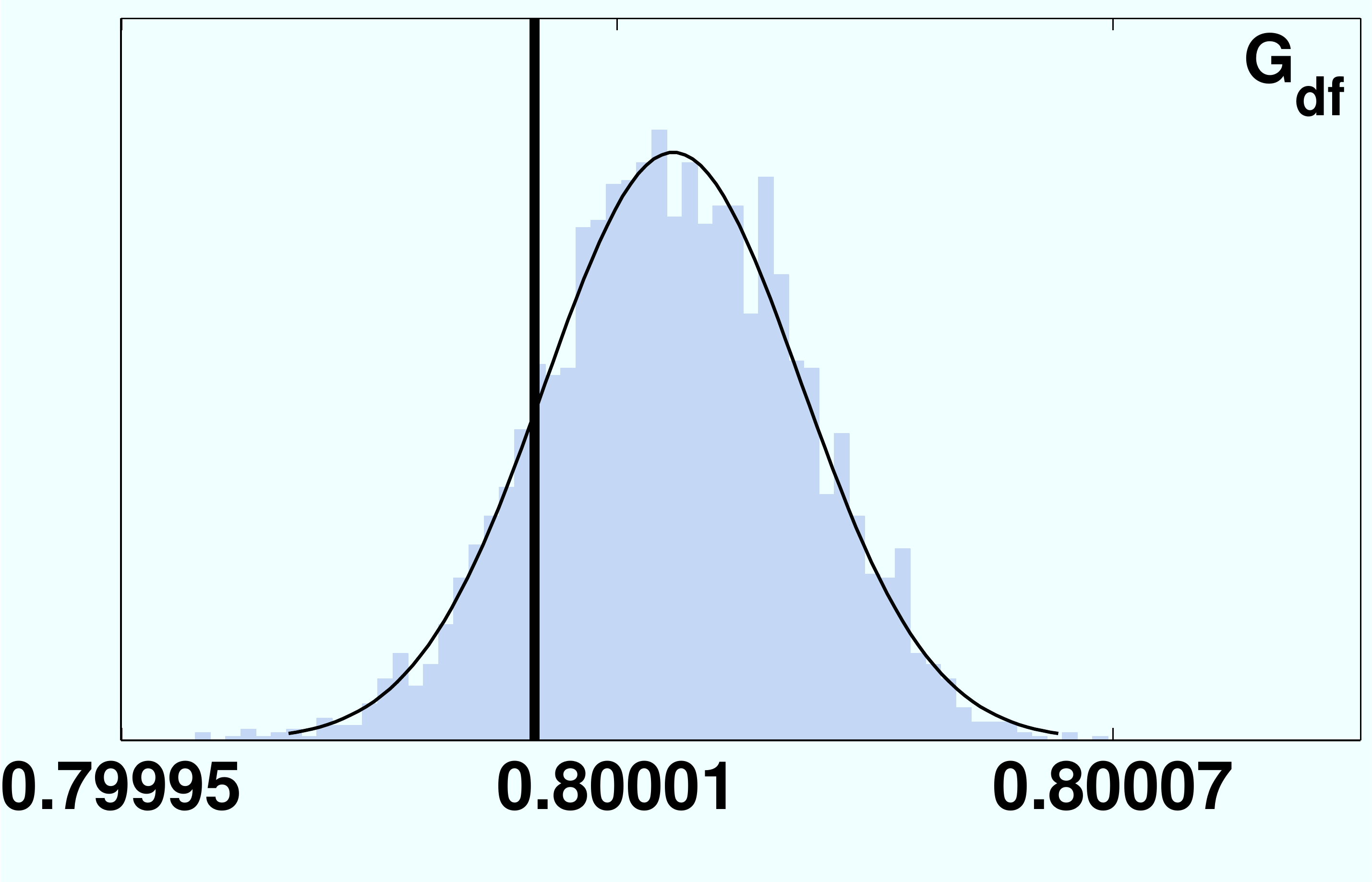} &   \includegraphics[scale = 0.19]{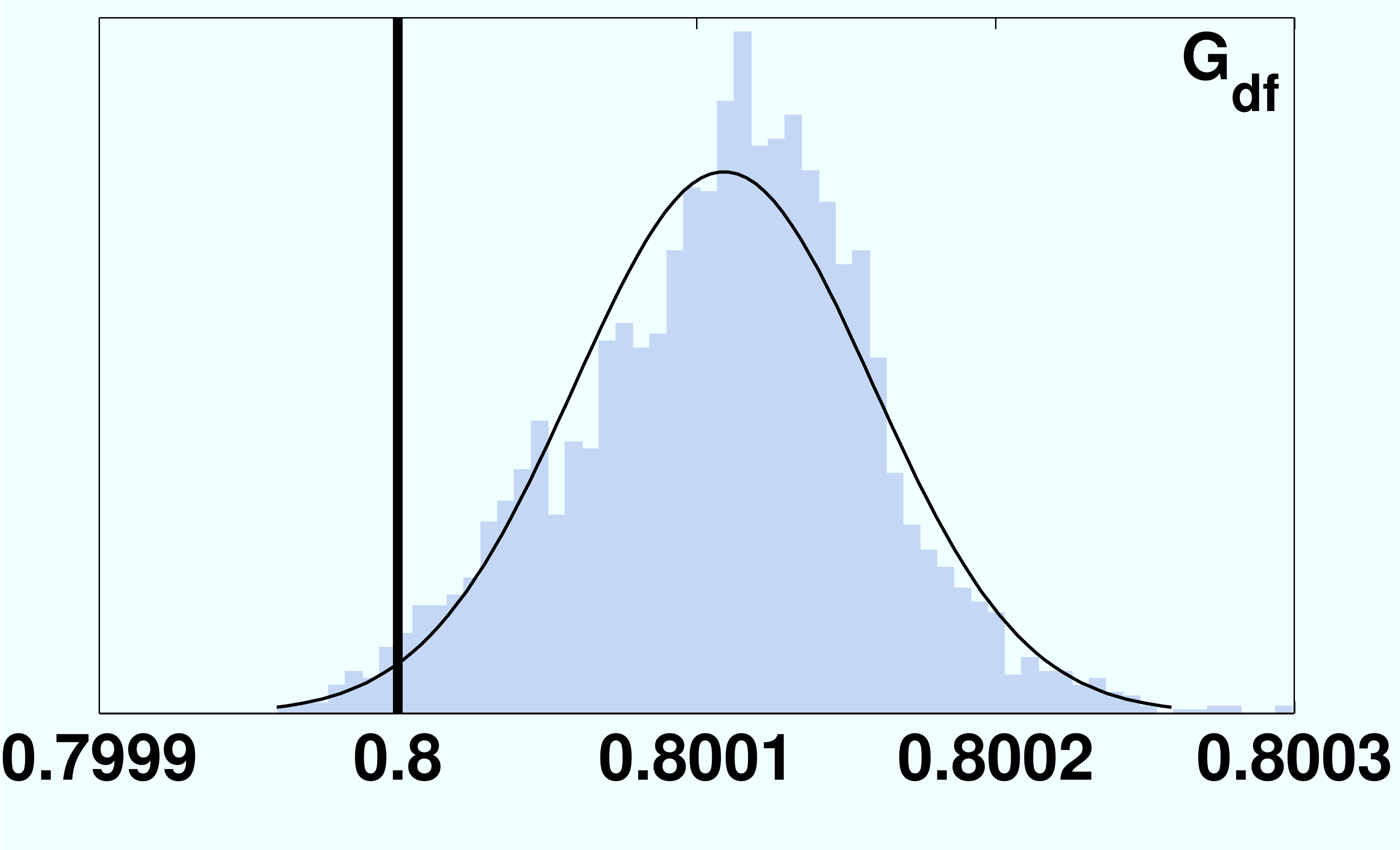} &  \includegraphics[scale = 0.19]{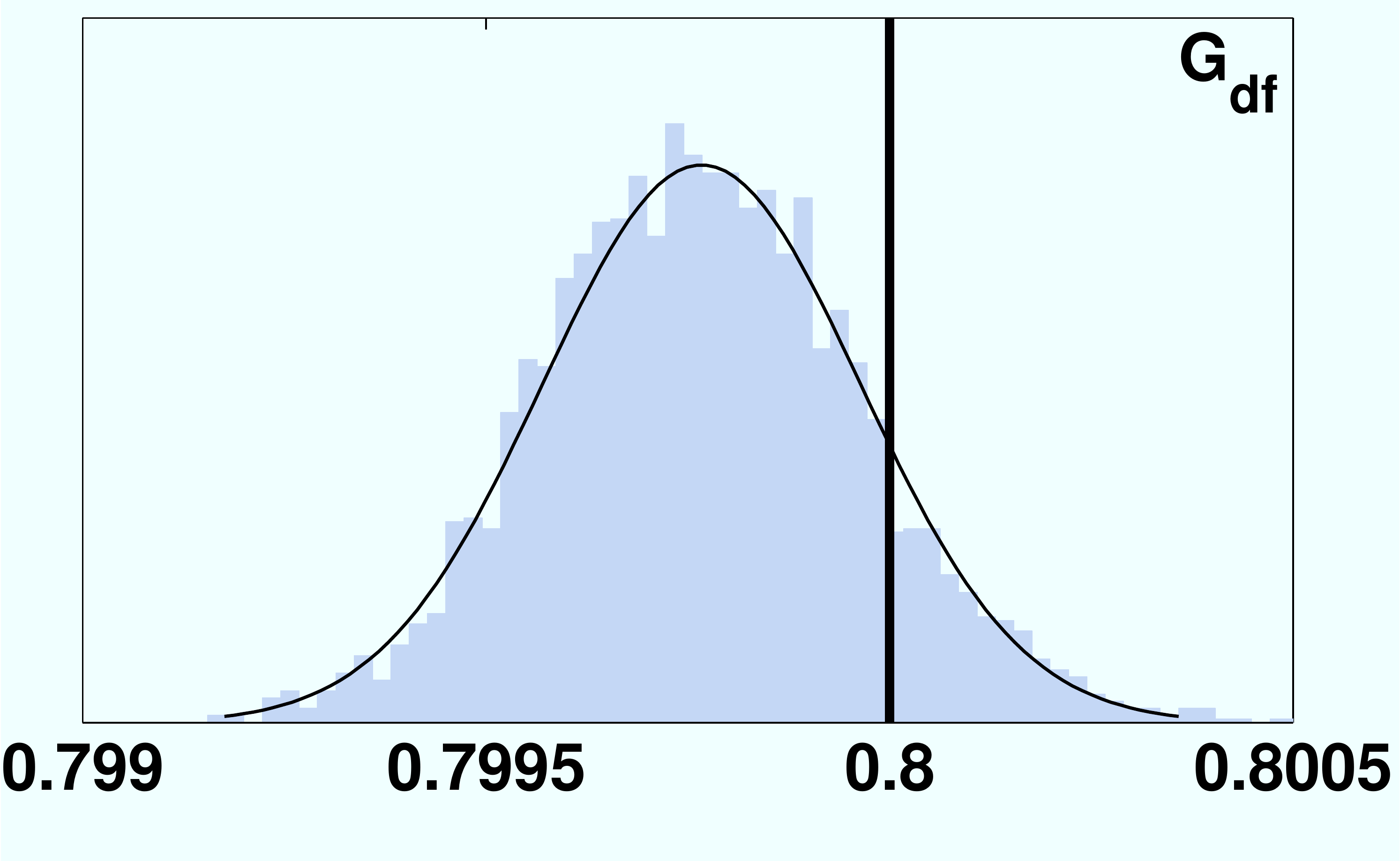} \\
 \includegraphics[scale = 0.19]{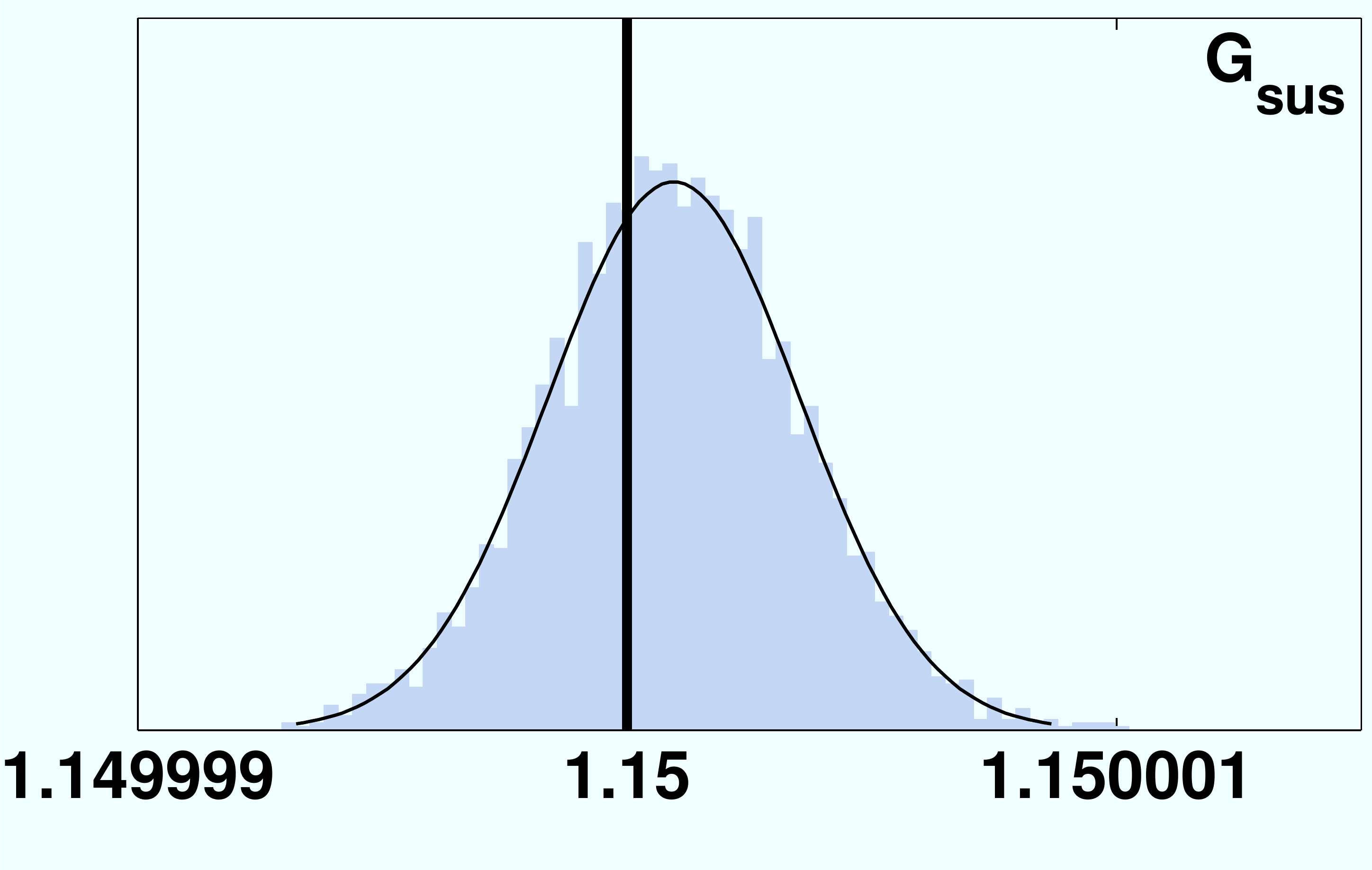} &   \includegraphics[scale = 0.19]{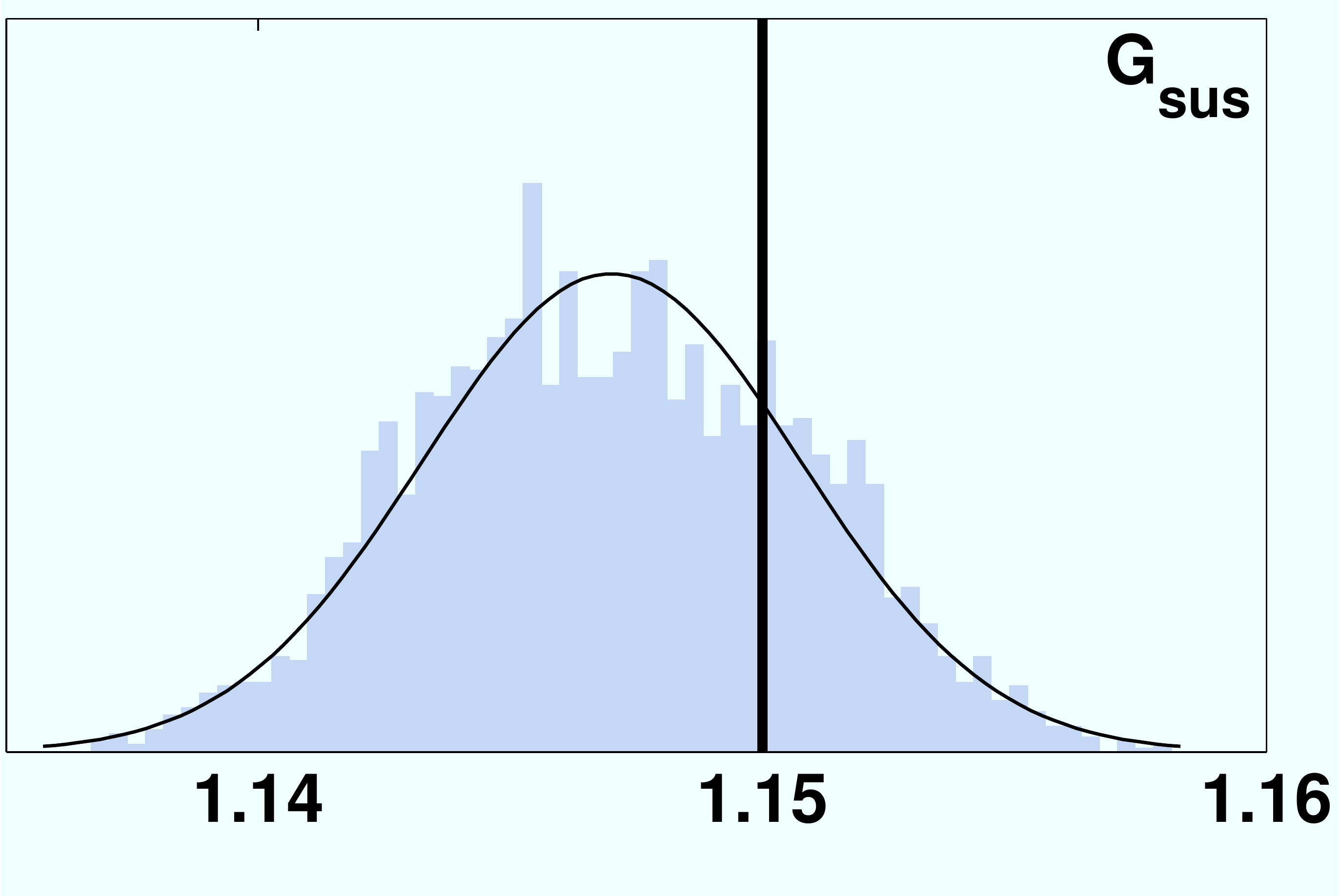} &  \includegraphics[scale = 0.19]{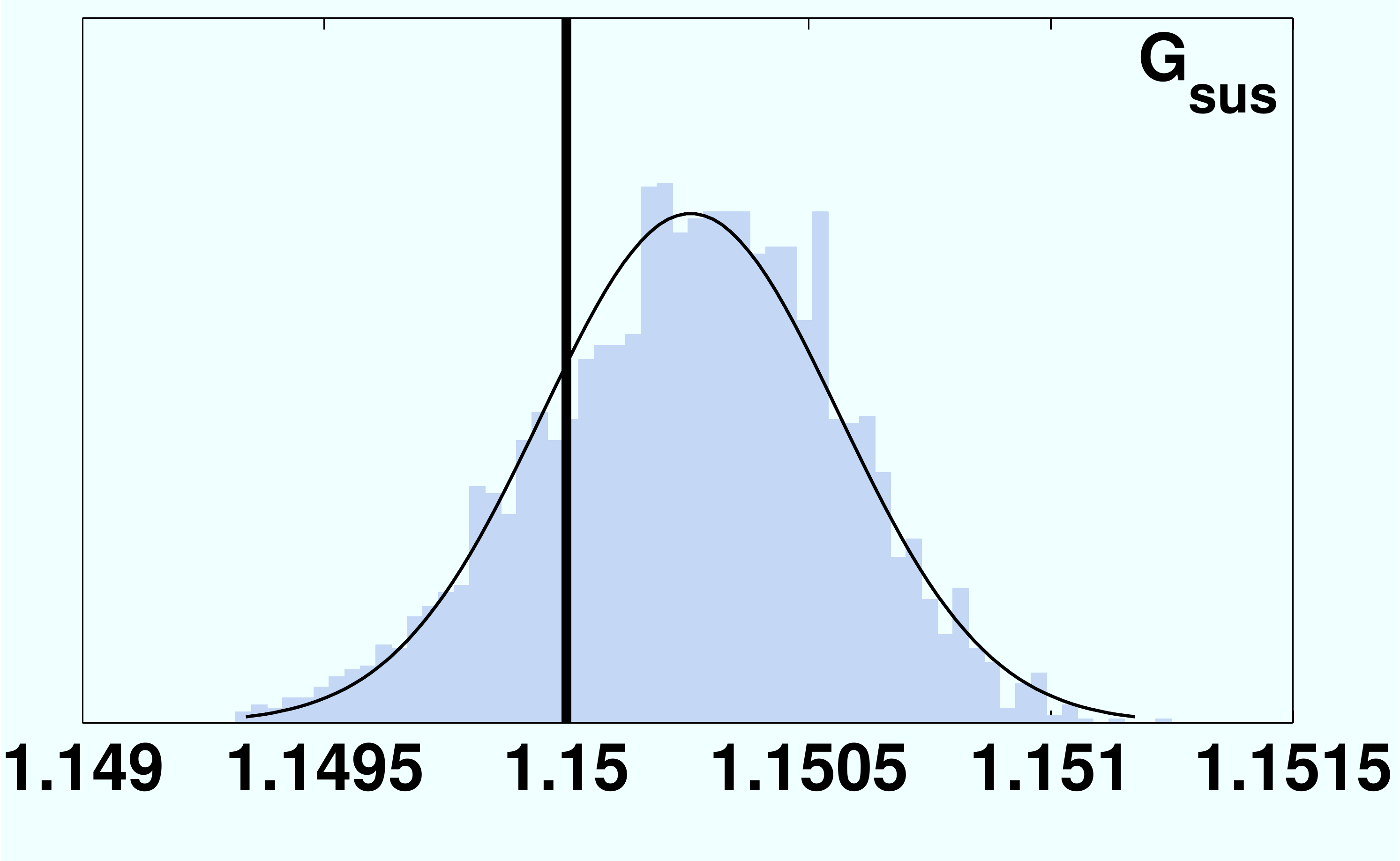} \\
 \includegraphics[scale = 0.19]{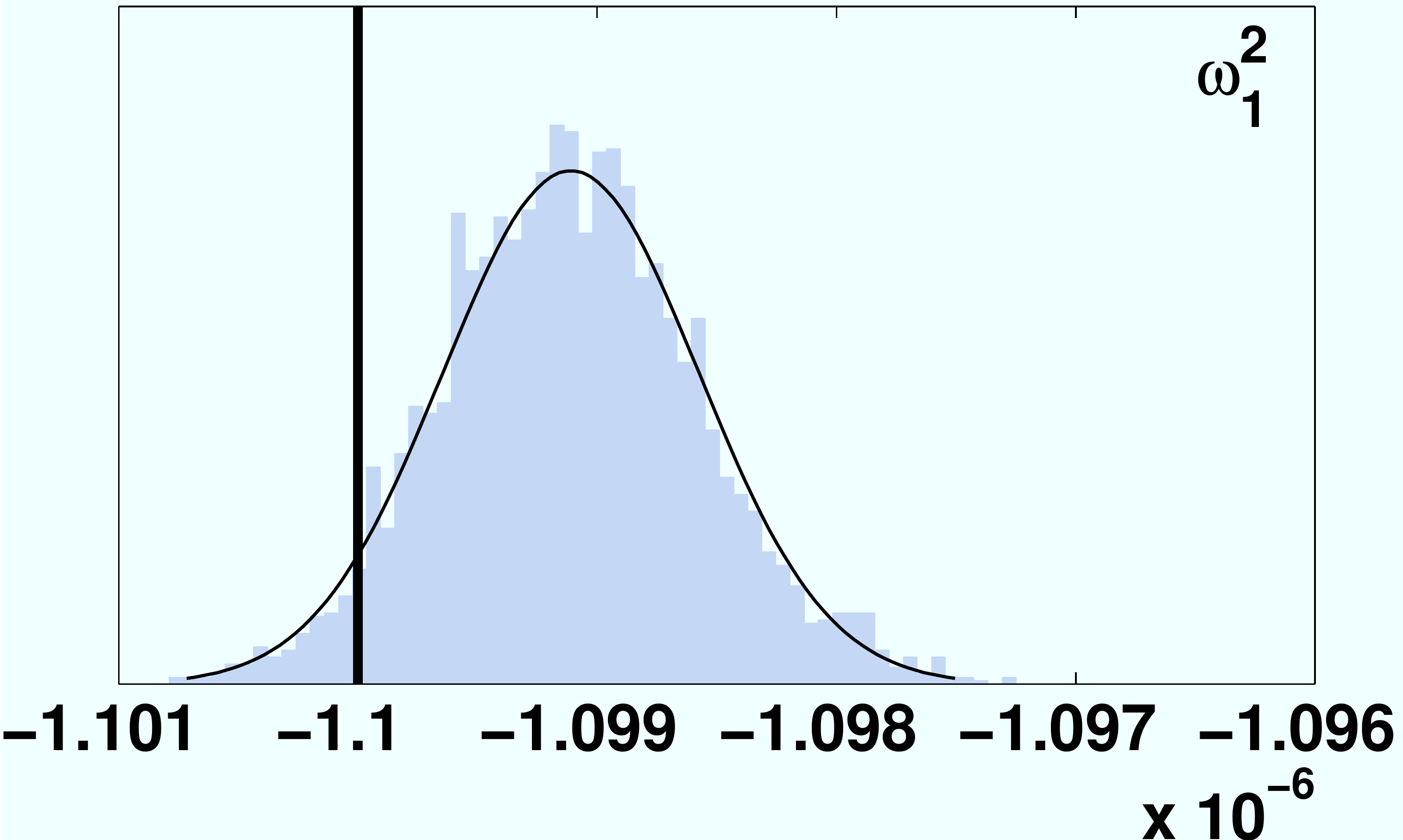} &   \includegraphics[scale = 0.19]{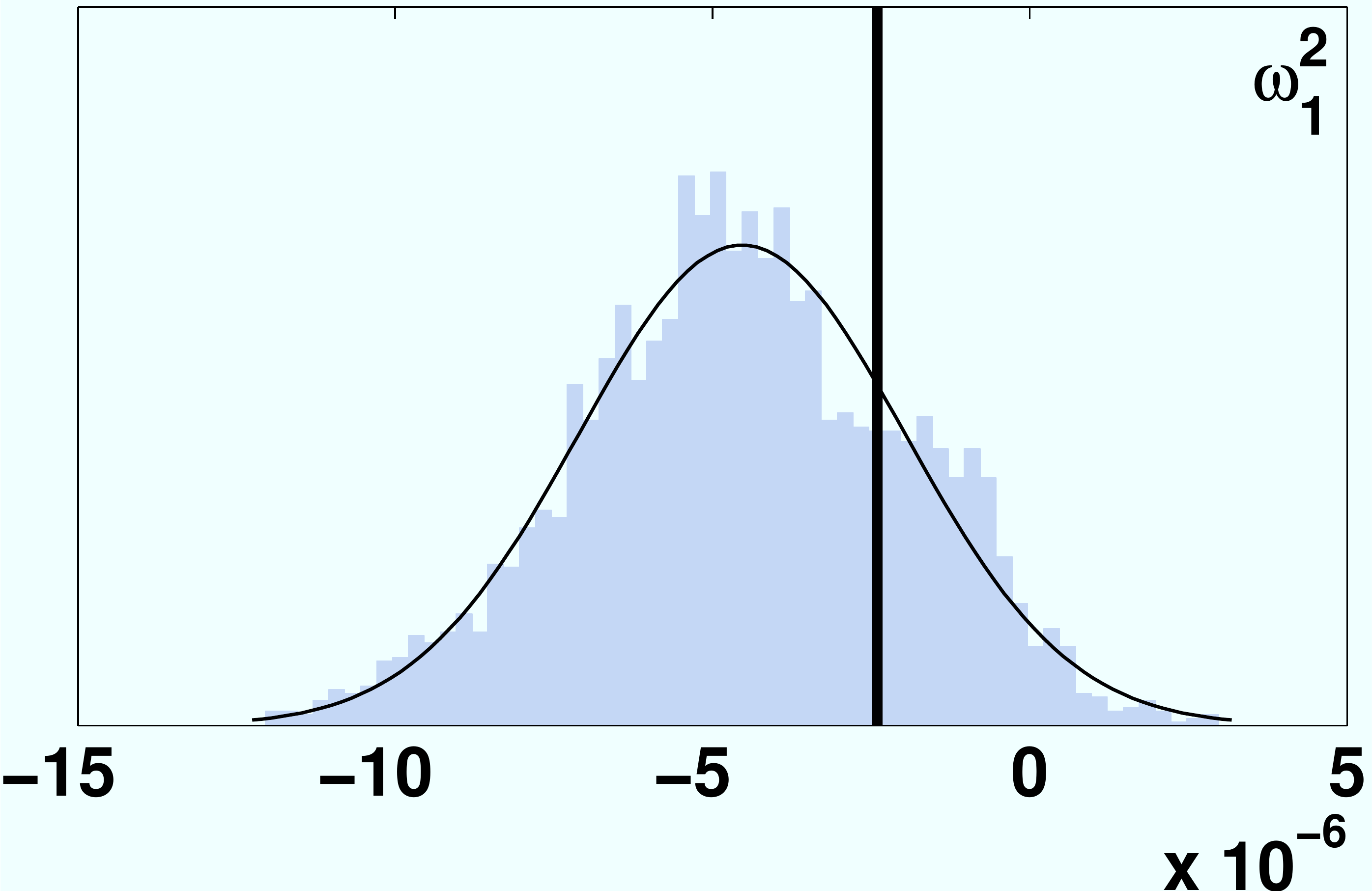} &  \includegraphics[scale = 0.19]{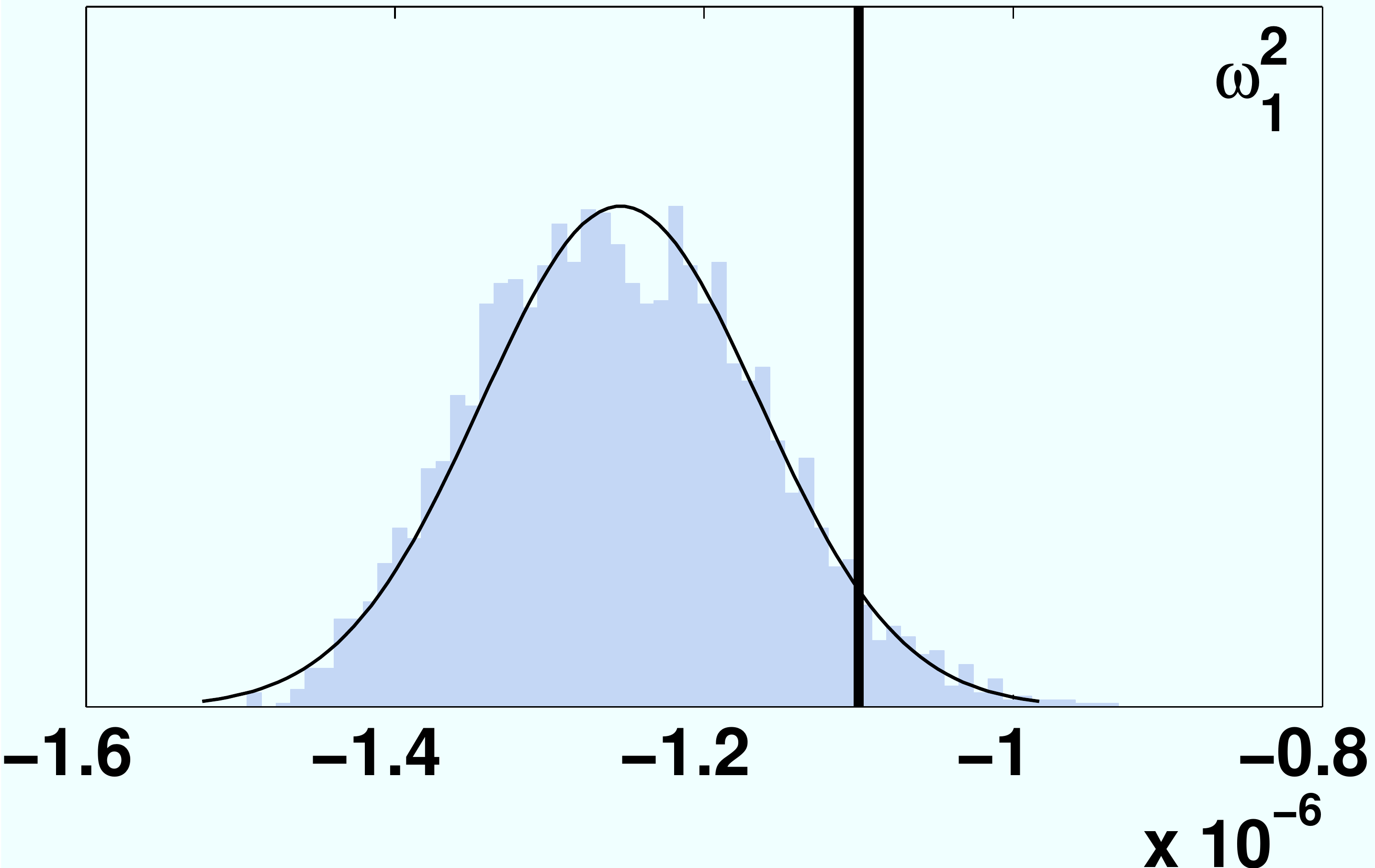} \\
 \includegraphics[scale = 0.19]{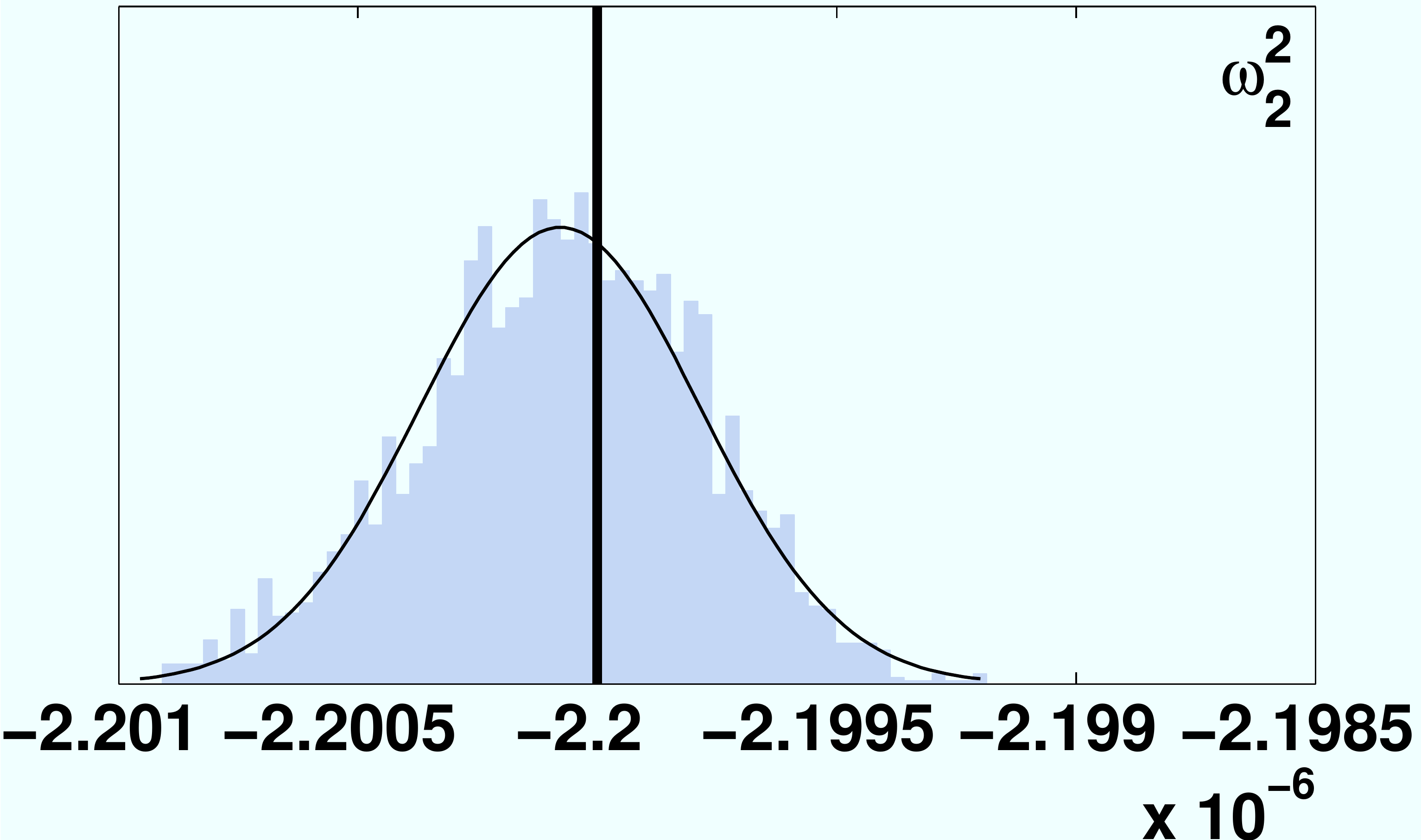} &   \includegraphics[scale = 0.19]{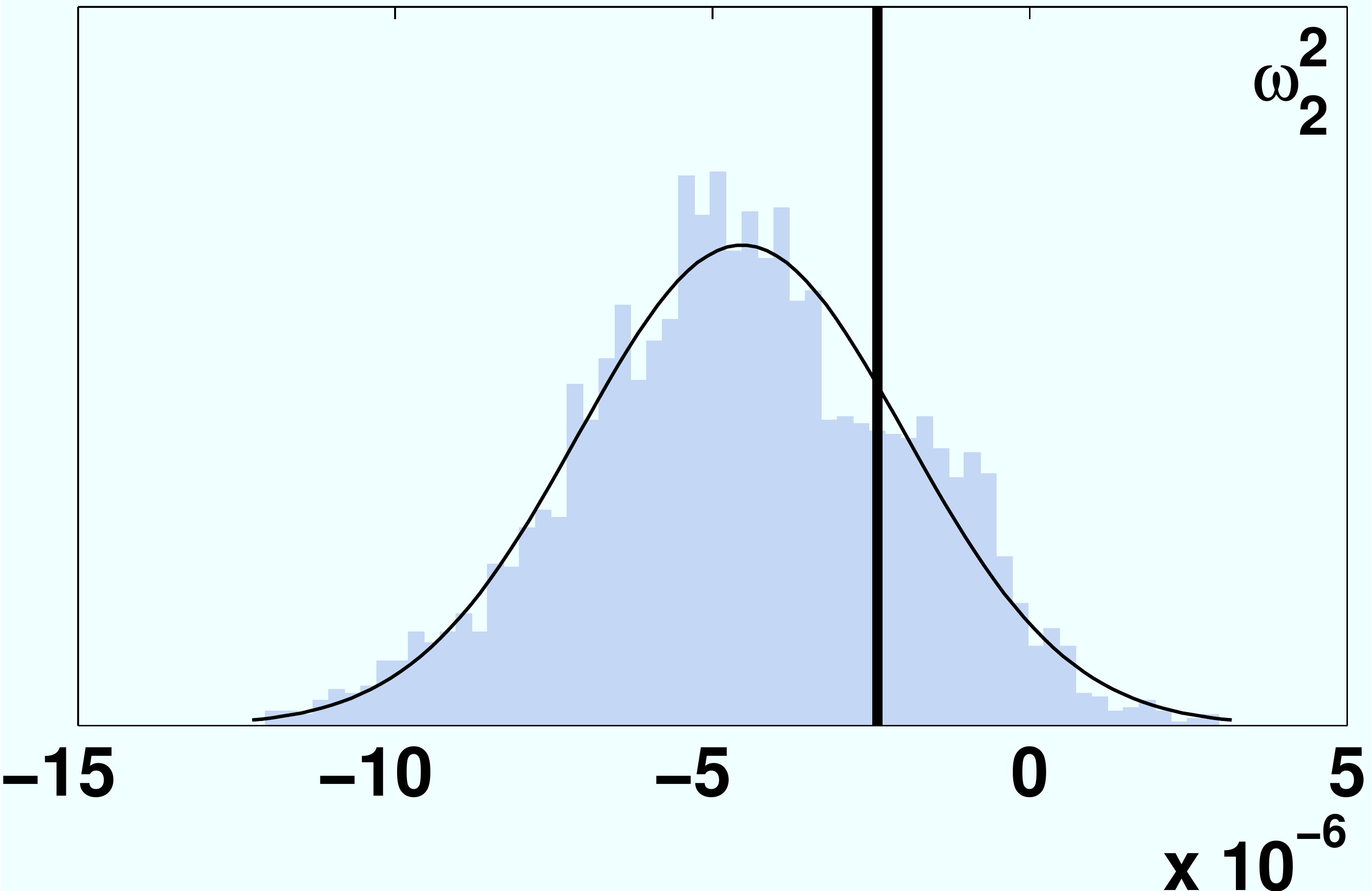} &  \includegraphics[scale = 0.19]{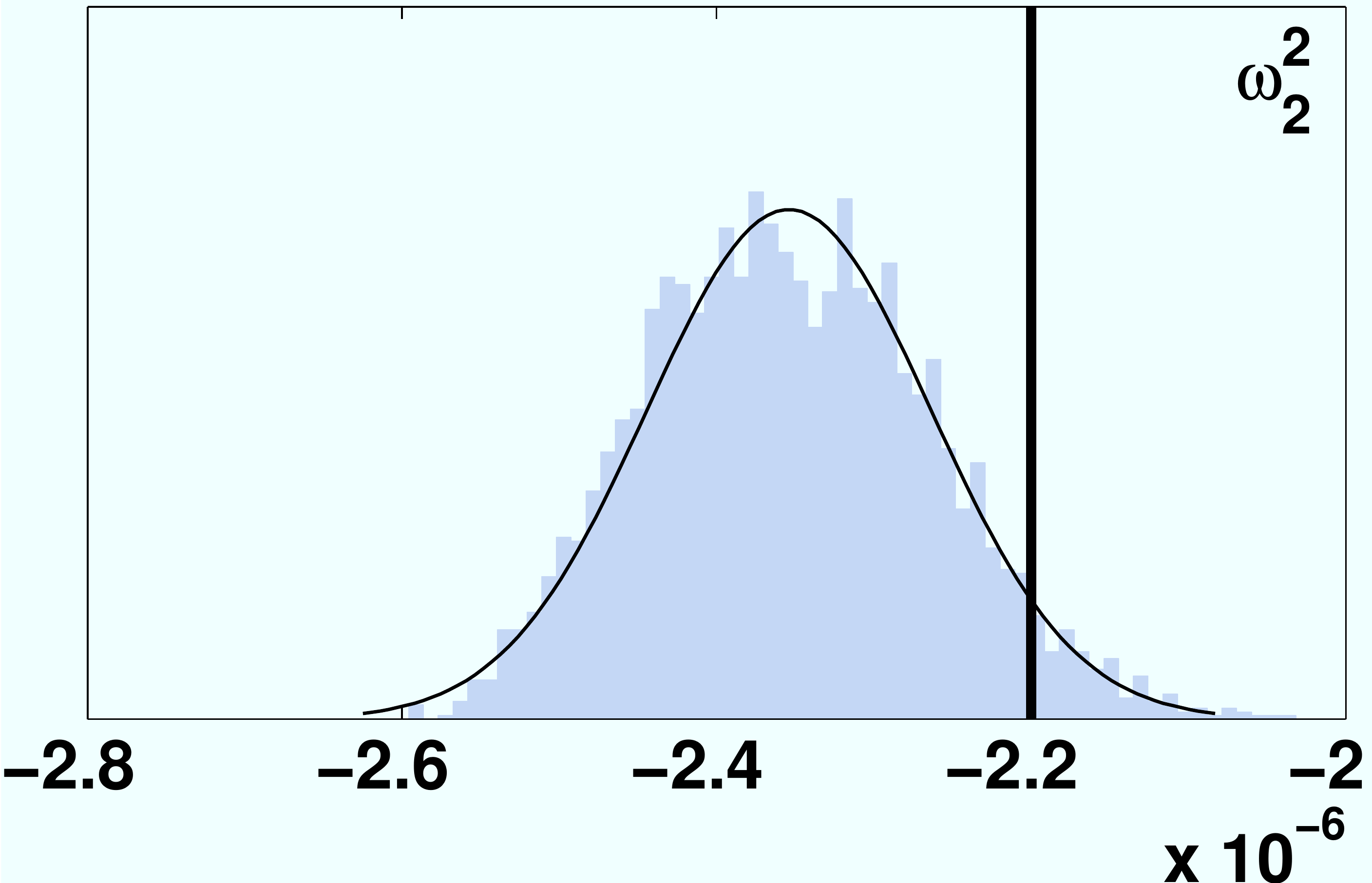} \\
 \includegraphics[scale = 0.19]{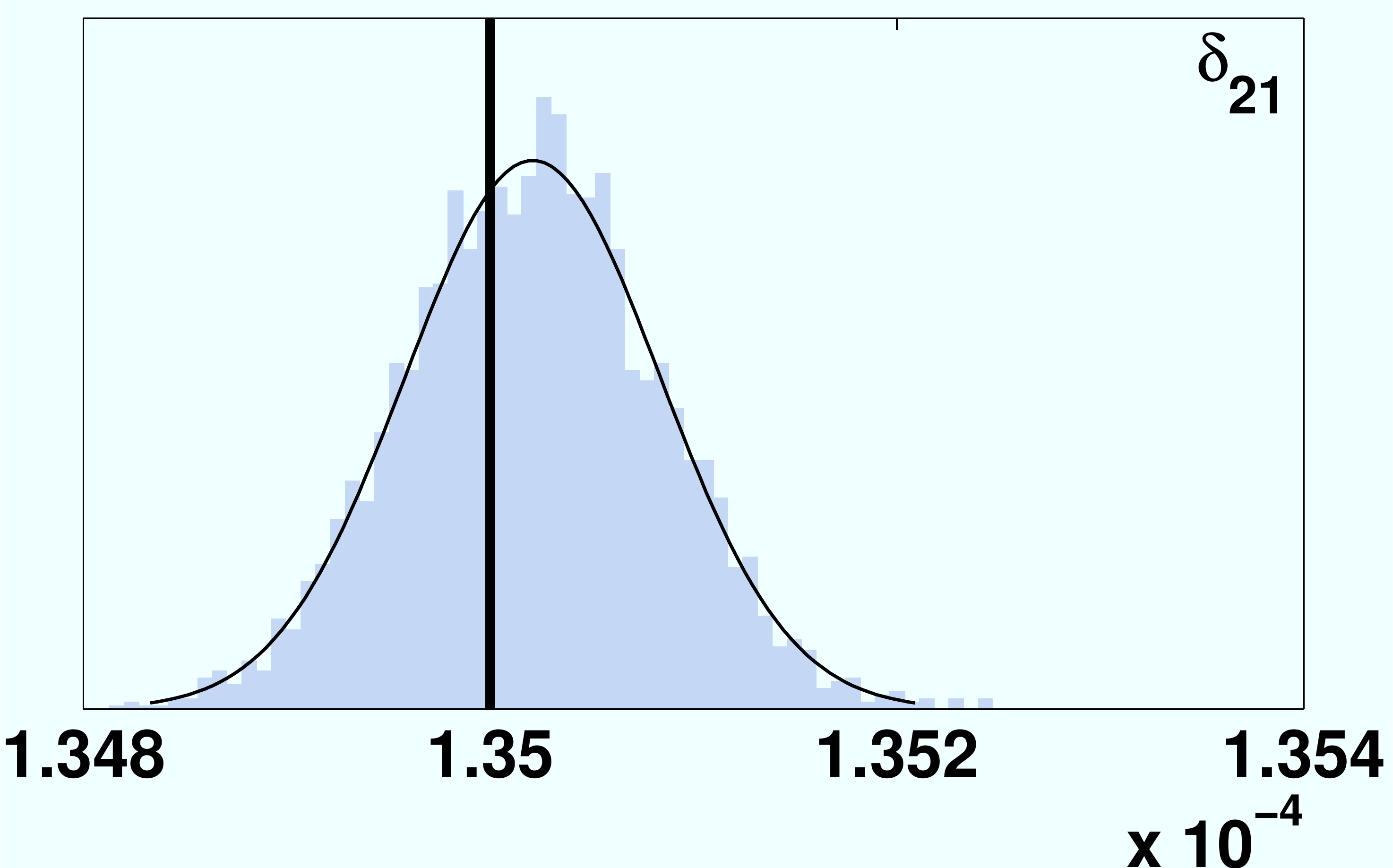} &   \includegraphics[scale = 0.19]{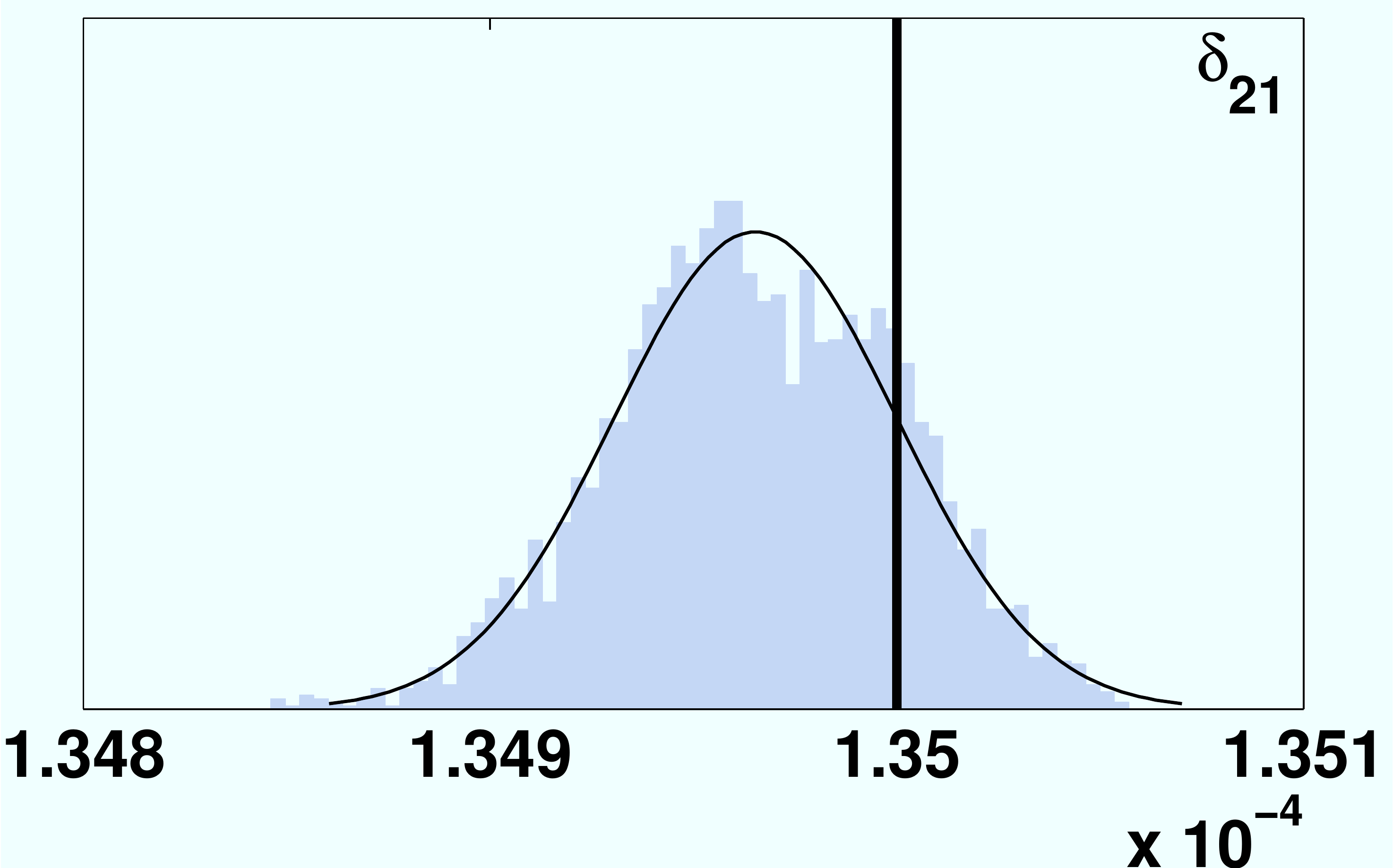} &  \includegraphics[scale = 0.19]{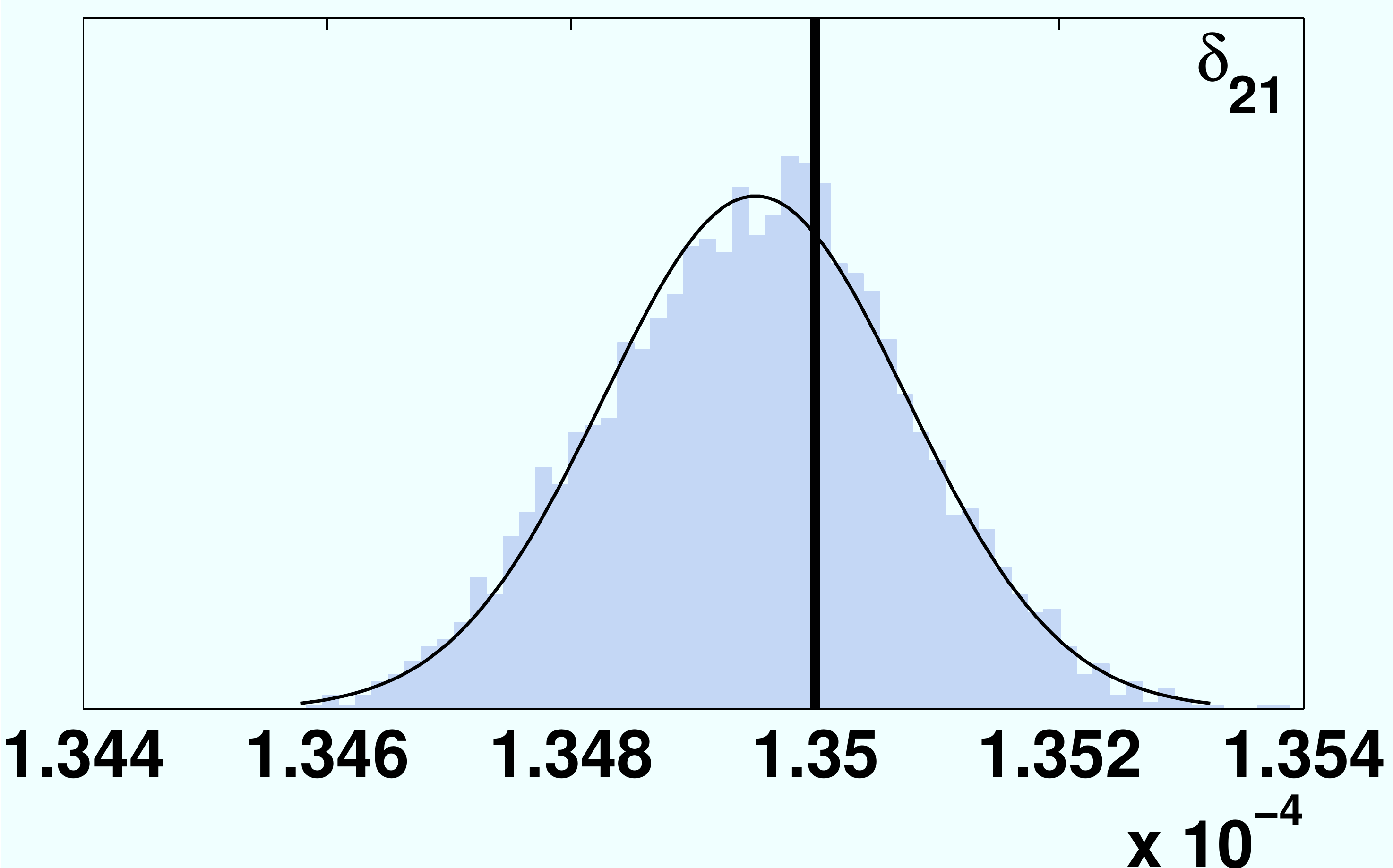} 
\end{array} $
\end{center}
\caption{Histograms of the MCMC samples illustrating the individual
 parameters' marginal posterior probability distributions as computed 
 with the last 3500 samples of the chain. 
 All histograms are plot with the same y axes range, up to 250 counts. 
 Black vertical lines illustrate the true parameter values. 
 Parameters $G_{\rm df}$, $G_{\rm sus}$
 and $\delta_{21}$ are dimensionless; dimensions for stiffness parameters are
 $\left[ \rm \omega^2_{1} \right] = \left[ \omega^2_{2} \right] = \rm s^{-2}$.
 \label{fig.Expshist}}
\end{figure*}

Figure~\ref{fig.Expshist} illustrates the marginal posterior
probability density functions of the individual parameters based on
the different experiments.  Parameter estimates are shown in
Table~\ref{tbl.Exps}, together with a comparison of the estimated
error and the Cram\'{e}r-Rao bounds, as derived in
Section~\ref{sec.crb}.  The parameters are recovered successfully with
estimation uncertainties roughly following the corresponding CRB, as
shown in the last column of table~\ref{tbl.Exps}.
The worse estimate appears to be a $\sim2 \sigma$ 
deviation on the $ G_{\rm df}$ parameter in experiment~2. 
This result is still consistent with the true value 
used to generate the data. However, to further 
investigate this feature we generated a new set of data
using the same tools and parameters. The analysis 
of the new data did not reproduce an offset estimate, 
whence we discarded a systematic bias on 
$ G_{\rm df}$ parameter in experiment~2. 

%{ Another interesting
%feature can be also recognised in both table and plots: the 
%worse estimate is a $\sim 3\,\sigma$ deviation
%$\rm G_{\rm sus}$ in experiment~2. As we 
%already showed in section \ref{sec.crb}, the discrepancy must be 
%assigned to the high correlation between these parameters and 
%the stiffnesses in an experiment where the latters can not be precisely 
%estimated. This same feature is observed in the histograms in 
%Figure~\ref{fig.Expshist} when comparing $\rm G_{\rm sus}$ 
%and $\omega^2_2$ for the three experiments. Only in
%experiment~1, with an injected signal in the second channel, 
%we are able to reduce the deviation in the estimate of $\omega^2_2$
%and consequently determine $\rm G_{\rm sus}$ precisely.}

As expected, the best estimates come from the first experiment since
the signal is richer in that case. The fact that a signal is
injected on both channels makes this experiment the most sensitive in
terms of the determination of the stiffness difference between both
test masses, reaching indeed the CRB, and obviously translating into a
better estimate for the remaining parameters. 

Only the second experiment allows a better estimation of one of the
parameters, $\delta_{21}$, since in this case we are canceling the
second cross-coupling term, $\rm \omega^2_{2} - \omega^2_{1}$, by
forcing stiffnesses from both test masses to have the same value. As
expected, the absolute value of the stiffness can not be determined
accurately in such a case. The reason being that the matched stiffness
configuration is precisely designed to make the experiment insensitive
to stiffness differences, which naturally turns into a poor estimation
of the parameter. It is however remarkable that, thanks to the
cross-variance terms, we can have a good determination of the
difference between the two stiffnesses, which should be identically
zero in this case. That's indeed the value retrieved by our analysis
with an uncertainty of $7\times 10^{-10}\,\rm s^{-2}$. 

It is worth comparing here the results obtained with the analysis to
measured quantities. Although the numerical values may differ, it
may be relevant to compare the uncertainties of the values in order to
check that our model is in quantitative agreement with experiments
being performed. To do so we take the stiffness as our figure of merit
since it has been extensively characterized in the torsion pendulum
facility~\cite{Carbone03}.  
Recent experiments in this facility report a remnant stiffness coupling the test masses to the surrounding GRS prototype of $(−2.5\, \pm\, 0.1) \times 10^{−9}$\,N/m~\cite{Cavalleri09}. When scaled by the mass of the LTP test masses (1.96\,kg) so to be expressed in terms of force per unit mass, these figure becomes $(−1.28\, \pm\, 0.05) \times 10^{−9}\,\rm s^{-2}$,
which could be compared to the
uncertainty in the estimation of the stiffness in our model, which
reaches $3 \times 10^{-10}\,\rm s^{-2}$ for the second test mass stiffness in
experiment~1. The simplified noise model that we used for the analysis
therefore seems to be consistent with the numbers coming from
experiments. Both numbers are, however, orders of magnitude below the
required remnant stiffness on board the satellite of $14 \times
10^{-7}\,\rm s^{-2}$~\cite{TopLevel}.

The data analysis during LISA Pathfinder operations will be strongly
conditioned by the operations schedule. In Section~\ref{sec.combine}
we describe how to exploit the posterior distribution in order to
combine results from different experiments.  We applied that scheme to
our results in order to produce a unique set of parameters for both
cases previously described: all parameters being identical (experiment
1 \& 3) and experiments with different numerical values of the
parameters (combining all experiments).  Given the approximately normal
distribution of the parameters that we get from the Monte Carlo
integration in Figure~\ref{fig.Expshist}, it is justified to apply the
Gaussian formalism that we introduced in Section~\ref{sec.combine}.
In particular, we just need to apply
Equation~(\ref{eqn.updated.posterior3}) to our set of experiments.
Results in Table~\ref{tbl.combine} show an improvement in the
uncertainty of the estimate. According to (\ref{eqn.bayes.combined}),
the same scheme could be obtained by considering the posterior
distribution of one experiment as a prior for the following one. This
would also improve the convergence time of the search, which could be
an important consideration during operations.

\section{Summary and future work}
\begin{table}
\caption{Combination of results for different experiments. Two values
   are reported when combining all experiments for parameters $\rm
   \omega^2_{1}$, $\rm \omega^2_{1}$ and $\rm \Delta \omega^2$. The
   top one is the result obtained by combining the values for
   experiment 1 \& 3, the bottom one corresponds to the matched
   stiffness experiment. \label{tbl.combine}}
\begin{ruledtabular}
\begin{tabular}{lc|c}
Parameter   & \multicolumn{2}{c}{Estimated}  \\
\hline
 & Experiment 1 \& 3& All experiments \\ 
\hline
$\rm G_{df}$ 					&  $0.800\,02 \pm 0.000\,02$              &  $0.800\,03 \pm 0.000\,01$  \\[0.3cm]
$\rm G_{sus}$  					&  $1.150\,000\,1 \pm 0.000\,000\,3 $     & $1.150\,000\,9 \pm 0.000\,000\,3 $\\[0.3cm]
$\rm \omega^2_{1} $   		&  $(-1.100\,0 \pm 0.000\,4)$  & $\begin{array}{c} (-1.100\,0 \pm 0.000\,4)  \\ (-5 \pm 3) \end{array}$ \\[-0.3cm]
$ (\times\,10^{-6})$ & & \\[0.3cm]
$\rm \omega^2_{2} $   		&   $(-2.200\,1 \pm 0.000\,3) $ 	& $\begin{array}{c}(-2.200\,0 \pm 0.000\,3)  \\ (-5 \pm 3) \end{array} $\\[-0.3cm]
$ (\times\,10^{-6})$ & & \\[0.3cm]
$\rm \delta_{12}$  			& $ (1.349\,8 \pm 0.000\,5) $  &  $ (1.349\,67 \pm 0.000\,02) $  \\
$ (\times\,10^{-4})$ & & \\[0.3cm]
$\rm \Delta \omega^2 $	& $(-1.100\,2 \pm 0.000\,2) $  &  $\begin{array}{c}(-1.1002 \pm 0.000\,2)\\ (-0.0003 \pm 0.0006)  \end{array} $ \\[-0.3cm]
$ (\times\,10^{-6})$ & & \\[0.3cm]
\end{tabular}
\end{ruledtabular}
\end{table}      

We have shown how a Markov chain Monte Carlo method can be used for
parameter estimation in the LISA Pathfinder mission.  In order to
demonstrate so, we generated data from a simplified model of the main
experiment on board the mission, the LTP\@.  This data set contains runs
where we injected signals to test the instrument, which must allow the
recovery of the parameters, and also some runs without any injection,
used to evaluate the noise performance of the instrument. We think
that the model used in our analysis serves as a complementary approach
to the already existing LISA simulators, since it includes some more
detail in the test mass dynamics and its coupling to the test mass
motion, precisely one of the key points that LISA Pathfinder aims to
investigate.

The analysis presented here includes an estimate of the 
optimal error achievable (for an unbiased estimate) 
for a given injected signal and
a configuration of the experiment. These results are of 
relevance for the mission since they show that it is as
important to develop data analysis tools as to to carefully design 
the experiment to be performed in flight. With our model, a different 
injection signal showed to improve two orders of magnitude
the estimation of the test mass stiffnesses --- results for 
experiment 1 and 3 in Table~\ref{tbl.crb}. Although the
expected parameter uncertainties in the real mission 
will be larger than the ones reported here, the dependencies 
on the parameters are representative. Thus, the 
decrease on the optimal error could be applicable to
the real mission as well. We will need however to confirm 
this result with more realistic models. 

The method developed here to analyse the data reaches
roughly the optimal 
attainable error for each single experiment. The combination 
of the results for different experiments obviously reduces the
uncertainty on the parameters, reaching lower errors than the 
ones originally derived from the Cram\'{e}r-Rao bound for each 
independent experiment.
When combining different experiments, our analysis took advantage 
of the gaussian posterior obtained during the sampling 
of the likelihood surface, so that a simple algebraic operation 
between gaussian distribution was enough to derive a combined
estimate of all experiments. However, the framework is general 
enough to include non-gaussian profiles, given that the full profile 
of the posterior is obtained during the sampling of the likelihood 
surface. 

The combination of estimates was performed here as an off-line
operation, i.e. after all experiments were analysed. A natural 
extension to this work would be to use the posterior distribution 
for a given experiment as prior for the next one, as motivated 
in Equation (\ref{eqn.bayes.combined}). This concept of
a chain of experiments is particularly suitable for LISA 
Pathfinder since, during flight operations,
we will naturally need to include results of previous experiments in
the next foreseen ones.
In other words, if the test mass stiffnesses are clearly 
determined in an experiment we may want to use that 
information for forthcoming experiments in order to 
effectively reduce the dimension of our problem. 
The method described here provides a way to 
include this information in the analysis in a clear way.
Moreover, the capability to use this information could be a powerful advantage
during operations due to the reduction of convergence time
that it implies.  

An increase in the uncertainty on the estimates is to be expected
when dealing with a more realistic model due to the increase
in dimensions of the parameter space. This is precisely the 
step that we will face in the forthcoming activities 
in preparation for the LTP data analysis. Our 
aim is to study in detail the experiments
defined to be implemented in flight, now that 
the basic functionality of the parameter estimation tool
is already demonstrated. In that sense, next steps 
will include a three dimensions
model and more complex injected signals, that will make use of the full 
capabilities of the spacecraft. This work is ongoing 
and will be presented in due time.
\\
\begin{acknowledgements}
We would like to thank Curt Cutler and Ed Porter for very fruitful
discussions in differents stages of this work. MN wants to thank a 
grant from Generalitat de Catalunya.
\end{acknowledgements}

%\input{include/appendix}
%^\input{include/tablefig}


\bibliographystyle{apsrev4-1}
\bibliography{library}

\end{document}